\documentclass[12pt]{article}
\usepackage{parskip}
\usepackage[margin=1in]{geometry}
\usepackage{setspace}
\usepackage{lineno}
\usepackage{amsmath,amssymb,amsfonts,amsthm}
\usepackage{mathtools}
\usepackage{graphicx}
\usepackage{booktabs}
\usepackage{multirow}
\usepackage{threeparttable}
\usepackage{siunitx}
\usepackage{subcaption}
\usepackage{float}
\usepackage{enumitem}
\usepackage{xcolor}
\usepackage{array}
\usepackage{rotating}

\usepackage{microtype}
\usepackage{comment}
\usepackage{ragged2e}
\usepackage{hyphenat}

\usepackage{tcolorbox}
\usepackage{mathrsfs}
\usepackage{calligra}
\usepackage{listings}

\allowdisplaybreaks

\usepackage[authoryear,round]{natbib}
\bibpunct{(}{)}{;}{a}{ }{,}

\usepackage[
    colorlinks=true,
    linkcolor=blue,
    citecolor=blue,
    urlcolor=blue
]{hyperref}
\usepackage{cleveref}

\makeatletter
\renewcommand{\NAT@open}{(}
\renewcommand{\NAT@close}{)}
\makeatother

\theoremstyle{plain}

\newtheorem{Th}{Theorem}

\theoremstyle{definition}

\theoremstyle{remark}

\def\var{\hbox{var}}
\def\cov{\hbox{cov}}

\def\wh{\widehat}

\def\0{{(0)}}

\def\mI{\mathcal{I}}
\def\mJ{\mathcal{J}}
\def\mK{\mathcal{K}}

\def\tr{{\rm trace}}
\def\trace{{\rm trace}}
\def\var{{\rm var}}

\DeclareMathAlphabet{\mathcalligra}{T1}{calligra}{m}{n}
\DeclareFontShape{T1}{calligra}{m}{n}{<->s*[2.2]callig15}{}
\newcommand{\scripty}[1]{\ensuremath{\mathcalligra{#1}}}

\def\boxit#1{\vbox{\hrule\hbox{\vrule\kern6pt
            \vbox{\kern6pt#1\kern6pt}\kern6pt\vrule}\hrule}}

\definecolor{codegreen}{rgb}{0,0.6,0}
\definecolor{codegray}{rgb}{0.5,0.5,0.5}
\definecolor{codepurple}{rgb}{0.58,0,0.82}
\definecolor{backcolour}{rgb}{0.95,0.95,0.92}

\makeatletter
\renewcommand{\maketitle}{%
  \begin{center}
    {\large\bfseries \@title\par}
    \vspace{2em}
    {\normalsize \@author\par}
  \end{center}
  \vspace{3em}
}
\makeatother

\title{\textbf{CARhy: Comprehensive Analyses of Circadian Rhythms in Transcriptomic Experiments with Multiple Conditions}}

\author{%
Weiyi Huang\textsuperscript{1,3}, 
Jerome S. Menet\textsuperscript{2},
Samiran Sinha\textsuperscript{1,*}\\[0.5em]
\textsuperscript{1} Department of Statistics, Texas A\&M University, College Station, Texas, USA\\
\textsuperscript{2} Department of Biology, Center for Biological Clock Research, Texas A\&M University, College Station, Texas, USA\\
\textsuperscript{3} Department of Biostatistics and Data Science, University of Texas Health Science Center at Houston, Houston, Texas, USA\\
Contact email: sinha@stat.tamu.edu 
}

\begin{document}
{\setstretch{2}\selectfont\maketitle}

\newpage

\section*{Abstract}

Circadian rhythms are endogenous oscillations that regulate various physiological processes and their disruption has been linked to many diseases, making it important to determine how gene-expression rhythms are altered across genotypes, treatments, or environmental exposures. Existing approaches for circadian transcriptomic analysis are often limited to pairwise comparisons or to a single aspect of rhythmic behavior, making them inadequate for comprehensive inference in multi-condition experimental designs. We propose CARhy (Comprehensive Analysis of Rhythmicity), a unified statistical framework for transcriptomic data collected under more than two conditions. Based on first-harmonic Fourier regression, CARhy provides formal tests for the presence of rhythmicity and for differences across conditions in rhythmicity, amplitude, phase, and baseline level. By allowing condition-specific variances and accommodating unbalanced designs, the framework remains reliable under heteroscedastic noise and realistic sampling constraints. Simulations show that CARhy controls type I error and false discovery rates well while achieving higher power than existing approaches in challenging settings. In mouse liver transcriptomic data, CARhy offers an interpretable and practical tool for characterizing how circadian rhythms differ across multiple experimental conditions. CARhy is implemented as an R package and is publicly available at: \url{https://github.com/DrHuang123/Comprehensive-Analyses-of-Circadian-Rhythms-CARhy.git}.

\textbf{Keywords:} Chi-square distribution, Circadian rhythms,  
Differential rhythmicity, F-test, Heteroskedasticity, 
Satterthwaite–Welch approximation.

\newpage

\section{Introduction} 

Circadian rhythm is an endogenous physiological process widely present in living organisms and characterized by 24-hours periodic changes. The disruption of this process is closely associated with the onset of multiple diseases, including metabolic disorders \citep{PMID39007272}, cardiovascular diseases \citep{PMID30796369}, and cancer \citep{PMID34615982}. Consequently, the factors that influence circadian rhythms and the molecular mechanisms involved have become an active focus of research in recent years. Circadian rhythms are found in almost all organisms and in most somatic cells throughout the body \citep{PMID31847894}. Among the available model systems, mouse liver \citep{PMID37382061} and Drosophila melanogaster \citep{PMID22472103} have been investigated most extensively. In mouse liver, a large number of transcripts exhibit significant rhythmic changes in abundance, thereby exerting important effects on cellular activities, energy metabolism, and diverse physiological processes \citep{PMID30995463, PMID35921362}. By comparing transcriptomic expression profiles across two or more conditions that differ in genotype, treatment regimen, or environmental exposure, researchers can in principle assess how external interventions or physiological perturbations reshape circadian rhythms and identify genes with differential rhythmic expression between conditions, thereby providing insight into underlying molecular mechanisms.

However, despite the growing availability of such multi-condition circadian transcriptomic studies, there is currently no unified statistical framework or dedicated software that provides a comprehensive, principled analysis of rhythmicity across multiple conditions while simultaneously disentangling distinct aspects of circadian behavior (such as changes in amplitude, phase, and overall rhythmic structure). Existing approaches are typically limited to pairwise comparisons or focus on individual components of rhythmicity in isolation, leaving a critical gap in the analysis of complex experimental designs.

To address this need, we propose a statistically principled and comprehensive framework for analyzing rhythmicity in such data, termed CARhy, where CA stands for \textit{Comprehensive Analysis} and Rhy stands for \textit{Rhythmicity}. This method is designed to fill the existing gap by enabling an integrated analysis of circadian patterns across multiple conditions within a single coherent modeling framework.

When analyzing circadian transcriptomic data, the key challenge is identifying which genes show evidence of differential rhythmic expression between experimental conditions. To address this, a variety of strategies have been employed. Early studies commonly used Venn diagram-based differential rhythmicity analysis \citep{PMID34189845}, in which rhythmic genes are first detected separately in each condition and then compared to identify gains or losses of rhythmicity; however, this procedure does not directly test the null hypothesis that rhythmic patterns are unchanged between conditions and tends to overestimate the number of differentially rhythmic genes. Subsequently, several dedicated statistical methods were developed, including parametric approaches for differential analysis of rhythmicity, such as DODR \citep{PMID27207944}, CircaCompare \citep{PMID31588519}, compareRhythms \citep{PMID34189845} and LimoRhyde \citep{PMID30472909}, as well as nonparametric approaches for detecting rhythmicity, such as JTK\_CYCLE \citep{PMID20876817} and RAIN \citep{PMID25326247}. These methods have improved statistical rigor and power to some extent, but important limitations remain. First, most are restricted to pairwise comparisons and provide the test for overall difference in rhythmicity, making it difficult to disentangle changes in amplitude and phase \citep{PMID27207944, PMID31588519}. Second, their performance can be sensitive to the number of sampling time points, the number of biological replicates, and the noise level \citep{PMID27207944, PMID20876817, PMID25326247}; as a result, different algorithms often yield markedly different counts of differentially rhythmic genes under the same statistical threshold, and type I error control is not always satisfactory.

\cite{PMID33452134} proposed a Bayesian model selection approach that flexibly accommodates more than two conditions for detecting differential rhythmicity of genes across multiple settings. Their framework bases inference on posterior model probabilities approximated using the Bayesian Information Criterion (BIC). While this strategy is powerful and general, it requires fitting a relatively large number of candidate models for each gene, which can lead to substantial computational cost. In addition, as is well known, the strong penalty term in BIC (proportional to the logarithm of the sample size) may favor simpler models, particularly when the true data-generating mechanism is not included among the candidate set.

We develop a general and computationally efficient method for testing differential rhythmicity, including separate tests for differences in amplitude and in phase across multiple conditions. Specifically, we model a gene’s  smoothly varying expression profile $\{Y(t)\}$, $t\in [0, 24]\}$
with a (first-order) harmonic regression.   

We fit the proposed model separately under each experimental condition and construct hypothesis-specific novel test statistics based on the estimated model parameters. The choice of the key components of the test statistics is guided by established principles of mathematical statistics.   
To obtain valid p-values, we derive the null (reference) distributions of these statistics  that verify differential effects, using the Satterthwaite-type approximation framework \citep{Satterthwaite1946, Welch1951}, which enables accurate inference without relying on equal-variance assumptions. The resulting p-values can then be  adjusted for multiplicity using established false discovery rate (FDR) control procedures \citep{Benjamini1995, Storey2003}.  For identifying rhythmicity under a condition, not only we use a novel test statistic but also a novel Monte Carlo method to assess its reference distribution.

From a statistical perspective, the key methodological contribution of this work is to eliminate the commonly imposed homoskedasticity assumption in circadian transcriptomic analysis. This extension is particularly important in practice, where variability can differ substantially across experimental conditions and sample sizes within each group are often limited. In addition, the proposed framework naturally accommodates unbalanced experimental designs, making it applicable to realistic biological studies in which equal allocation across conditions is rarely achieved. The methodology is computationally efficient and is designed to scale to high-dimensional transcriptomic settings. Across extensive simulation studies, the proposed procedure maintains type-I error rates close to their nominal levels, achieves strong statistical power, and provides reliable false discovery rate control under a wide range of challenging scenarios.

From an applied perspective, the framework provides a unified and statistically principled toolkit for circadian rhythm analysis. It enables formal testing for the presence of rhythmicity and supports targeted, hypothesis-driven inference on differential rhythmic characteristics across conditions, including changes in mesor, amplitude, and phase. Importantly, in settings characterized by limited sample sizes—a common constraint in circadian experiments—we carefully construct the test statistics and their key components to ensure stable and reliable inference, addressing an aspect that has received limited attention in the existing literature. To the best of our knowledge, no existing methodology provides a unified framework that simultaneously supports these multiple layers of differential rhythmic inference while maintaining rigorous statistical guarantees under practical design constraints. To promote reproducibility and broad accessibility, we have implemented the proposed methodology in an open-source R package.

\section{Methods}

\subsection{Experimental Data Oriented Models and Assumptions}\label{sec2:maa}

We develop our modeling framework in the context of a motivating RNA-seq experiment \citep{PMID30995463}, while keeping the formulation sufficiently general to accommodate a broad class of circadian transcriptomic studies. In this experiment, 54 wild-type mice maintained under a standard 12-hour light/12-hour dark cycle were assigned to one of three feeding regimens: nighttime feeding (NF), ad libitum feeding (AL), and arrhythmic feeding (AF). Liver tissues were collected at six Zeitgeber times (ZT02, ZT06, ZT10, ZT14, ZT18, ZT22), corresponding to 2, 6, 10, 14, 18, and 22 hours after lights on, respectively, with three biological replicates per time point within each feeding condition.

Within this setting, our analysis intends  to address  following statistical objectives: (i) to identify genes that exhibit significant rhythmicity within each experimental condition; (ii) to detect genes that display differential rhythmicity across conditions; (iii) to identify genes with differences in mesor (baseline expression level) across conditions; (iv) to detect genes with differential amplitudes, reflecting changes in the magnitude of oscillation; and (v) to identify genes with differential phases, corresponding to shifts in the timing of peak expression. Collectively, these objectives enable a comprehensive characterization of how experimental conditions modulate circadian gene expression dynamics.

Motivated by this design, but not restricted to it, we adopt a general notation to describe multi-condition circadian experiments. Let $Y_{g,j,k}(t)$ denote the appropriately normalized, log$_2$-transformed, and batch-adjusted expression level of the $g$th gene measured from the $j$th sample at circadian time $t$ under condition $k$. We use $g=1,\dots,\mathcal{G}$ to index genes and $k=1,\dots,\mathcal{K}$ to index experimental conditions. The sampling times (Zeitgeber times) may differ across conditions; accordingly, for condition $k$, let $0 < t_{1,k} < \cdots < t_{\mathcal{I}_k,k} < 24$ denote the observed time points. Furthermore, the number of biological replicates at each time point may vary, so that for time $t_{i,k}$ under condition $k$, samples are indexed by $j=1,\dots,\mathcal{J}_{i,k}$.

This formulation captures the key features of the motivating experiment—multiple conditions, discrete sampling times, and limited replicates—while remaining flexible enough to accommodate unbalanced designs and heterogeneous sampling schemes commonly encountered in circadian studies.

We assume the following model for the expression of a given gene (dropped the suffix $g$) under condition $k$ based on the Fourier basis expansion of order one, 
\begin{align}\label{eq:1}
Y_{j, k}(t_{i, k})=\delta_{k}+\alpha_{k}
\cos\left(2\pi \frac{t_{i, k}}{24}\right)+\beta_{k} \sin\left(2\pi \frac{t_{i, k}}{24}\right)+\epsilon_{i,  j, k}, 
\end{align}

where the noise $\{\epsilon_{i, j, k},  j=1,\dots, \mJ_{i, k}, 
 i=1, \dots, \mI_k
 \}$ are assumed to be independent and identically distributed (iid) normal variables with zero mean and variance $\sigma^2_{k}$. 
Here $\delta_k$ denotes  rhythm adjusted mean for a 24h period, called mesor.   
The  amplitude representing the distance between the mesor to the peak, is $A_k=\sqrt{\alpha^2_k+\beta^2_k}$ and the phase, where the peak happens,  is $\phi_k
={\rm atan2}\left(\beta_k, \alpha_k\right)$, which carries the quadrant information. This phase is a measure of the angle in radians and it lies in $[-\pi, \pi]$.  
    Since the frequency is fixed, the  rhythmicity depicting amplitude and phase,  is fully characterized by 
$(\alpha_{k}, \beta_{k})$. 
Now, using model  (\ref{eq:1}), all expressions on the given mRNA from condition $k$ can be compactly expressed as  
\begin{align}\label{eq:mat}
Y_{k}= X_k\gamma_{k}+ \epsilon_{k},
\end{align}
where both the observed data $Y_k$ and unobserved noise $\epsilon_k$ are vectors of length 
$n_k=\mJ_{1, k}+\cdots+ \mJ_{\mI_k, k}$,  which represents the sample size for the $k$th condition and $X_k$ is the design matrix of order $n_k\times 3$ and $\gamma_k=(\delta_k, \alpha_k, \beta_k)^\top$. 
For example, we can set  
$Y_{k}=(Y_{1, k}(t_{1, k}), \dots,
Y_{\mJ_{1, k}, k}(t_{1, k}), \dots, 
Y_{1, k}(t_{\mI_k, k}), 
\dots, $ $Y_{\mJ_{\mI_k, k}, k}(t_{\mI_k, k}))^\top$,
then the row of $X_k$ corresponding to a component of $Y_k$ will consists of one and the allied harmonic terms. 
We assume, for any $k$, ${\rm rank}(X_k)=3$  and $\epsilon_k|X_k\sim {\rm Normal}(0, \sigma^2_kI_{n_k})$, where $I_{n_k}$ denotes the identity matrix of order $n_k$. 

The least square estimator of $\gamma_{k}$ is  $\widehat\gamma_{k}= (X^\top_k X_k)^{-1}X^\top_kY_{k}$, and the estimated covariance matrix of $\widehat
\gamma_{k}$
is 
$\wh\Sigma_{k}=\wh\sigma^2_{k}(X^\top_k X_k)^{-1}$,
$\wh\sigma^2_{k}= (Y_{k}- X_k\wh\gamma_{k})^{\otimes 2}/(n_k-3)$, where $a^{\otimes 2}=a^\top a$, for any generic vector $a$. 
Let $\Gamma=(\gamma^\top_{1}, \dots, \gamma^\top_{
\mK})^\top$
and 
$\wh\Gamma=(\wh\gamma^\top_{1}, \dots, \wh\gamma^\top_{\mK})^\top$. 
Denote the estimated covariance of $\wh\Gamma$ by $\wh\Sigma$ which is  
${\rm Diag}(\wh\Sigma_{1}, \dots, 
\wh\Sigma_{\mK})$.

\subsection{Inference Methodology}

\subsubsection{Test for differential rhythmicity (CARhy-TDR)}\label{AprxF}

To test if the rhythmicity of the gene varies across the conditions (e.g., varies across the feeding patterns for the wild-type mice), we test   
 $H_0: (\alpha_{1}, \beta_1)=\cdots=(\alpha_{\mK}, \beta_{\mK})$ versus 
$H_a:$ at least one of  $(\alpha_k, \beta_K)$-parameter vector is different from the rest.  
Define the $2(\mK-1)\times 3\mK$ contrast matrix
\begin{eqnarray*}
L=\left(
\begin{array}{rrr rrr rrr rrr}
0 & 1 & 0 &   & 0 & -1 & 0 &   & \cdots & 0 & 0 & 0 \\
\vdots \\
0 & 1 & 0 &   & 0 & 0 & 0 &   & \cdots & 0 & -1 & 0 \\
0 & 0 & 1 &   & 0 & 0 & -1 &   & \cdots & 0 & 0 & 0 \\
\vdots \\
0 & 0 & 1 &   & 0 & 0 & 0 &   & \cdots & 0 & 0 & -1 \\
\end{array}
\right)    
\end{eqnarray*}

and the hypothesis can be expressed 
as $$H_0: L\Gamma=0\mbox{ versus } L\Gamma\ne 0.$$
Our proposed test statistic for assessing differential rhythmicity (DR) is given by  
\begin{equation*}
T_{DR}= \frac{1}{\rho}(L\hat{\Gamma})^\top (L \hat{\Sigma} L^\top)^{-1} (L \hat{\Gamma}),
\end{equation*}
where \(\rho\) denotes the rank of \(L\). A standard but naive approach would be to approximate \(\rho T_{DR}\) by a \(\chi^2_\rho\) distribution. However, this approximation relies on large-sample asymptotics and is generally unreliable in the moderate to small-sample regimes that are typical of in vivo circadian experiments. For instance, in our motivating dataset there are six time points with three biological replicates per time point, resulting in a total sample size of only \(n_k = 18\). In such settings, the sampling variability of the estimated quantities can be substantial, and the resulting \(\chi^2\)-based approximation may lead to non-negligible distortions in inference. Moreover, $T_{DR}$ does not follow a straightforward $F$ distribution as the homogeneity assumption on the  variances is not adopted.  

A central contribution of this work is that we avoid this reliance on asymptotic arguments by deriving the exact finite-sample null distribution of \(T_{DR}\). This result provides a fully justified reference distribution that is valid under the sampling regimes encountered in practical circadian studies, thereby enabling accurate and principled inference without requiring large-sample approximations.

\begin{Th}\label{th1}
Under the stated assumptions, the distribution of a constant $c$ times $T_{DR}$ is approximately $F$ with the numerator degree of freedom $\rho$ and the denominator degree of freedom $df$, where 
\begin{eqnarray*}
    df=\frac{\rho-2 +2\rho \mu_2/\mu^2_1}{\rho\mu_2/2\mu^2_1-1}
\, \, and \, \, 
    c= \frac{\rho df}{\mu_1(df-2)}
\end{eqnarray*}
with
\begin{eqnarray*}
\mu_1= \rho+ \sum^\mK_{k=1}\frac{2\wh\sigma^4_{k}}{(n_k-3)}
\tr(B_k B_k),
\end{eqnarray*}
\begin{eqnarray*}
 \mu_2
&=&2\rho+\sum^\mK_{k=1} \var(\wh\sigma^2_{k})\biggl[
\{\tr(B_k)\}^2+ 6\tr(B_kB_k)\biggl]\\
 &&-\sum^\mK_{k=1}
E\{(\wh\sigma^2_{k}-\sigma^2_{k})^3\}\biggl
\{2\tr(B_k)\tr(B_kB_k)   +4 \tr(B_kB_kB_k)\biggl\}
\\
&&+  \sum^\mK_{k=1}
E\{(\wh\sigma^2_{k}-\sigma^2_{k})^4\}\biggl[
\{\tr(B_kB_k)\}^2   +2 \tr(B_kB_kB_kB_k)\biggl]\\
&&-\sum^\mK_{k=1} \var^2(\wh\sigma^2_{k})\{\tr(B_kB_k)\}^2\\
&& +\sum_{s>k} \var(\wh\sigma^2_k)\var(\wh\sigma^2_s)\biggl[
\{\tr(B_kB_s+B_sB_k)\}^2+4\tr(B_kB_kB_sB_s)\\
&&\hskip 5mm + 
 2\tr\{(B_kB_s+B_sB_k)(B_kB_s+B_sB_k)\biggl],
\end{eqnarray*}
$$
B_{1}=  L{\rm Diag}((X^\top_1 X_1)^{-1}, \cdots, 0) L^\top  
\Omega^{-1}, \cdots, 
B_{\mK}=  L{\rm Diag}(0, \cdots,  (X^\top_\mK X_\mK)^{-1}) L^\top \Omega^{-1}, 
$$
and $\Omega= L \Sigma L^\top$. 
\end{Th}
The proof of this result is given in the Supplementary Appendix~\ref{app:proof-thm1}. This result helps us calculating the p-value of the test. Specifically, for a gene, the p-value is then 
estimated by $P(F_{\rho, df}>cT_{DR})$, where $F_{\rho, d_2}$ denotes the $F$ random variable with degrees of freedom 
$(\rho, df)$, $df$ and $c$
are calculated using the formula given in Theorem \ref{th1}, and $T_{DR}$ is the observed value of the test statistic. 

Now, we discuss how to compute expression $\mu_2$ of Theorem \ref{th1} that involves several moments of $\wh\sigma^2_k$, $k=1, \cdots, \mK$ and $\Omega=L\Sigma L^\top$. 
First, we replace $\Sigma$ by $\wh\Sigma$ in $\Omega$. 
Next, since $\wh\sigma^2_{k}\sim \sigma^2_{k}(n_k-3)^{-1}\chi^2_{n_k-3}$, 
$\var(\wh\sigma^2_{k})=\sigma^4_{k}(n_k-3)^{-2}\times 2 (n_k-3)=2\sigma^4_{k}(n_k-3)^{-1}$, 
$E(\wh\sigma^2_{k}-\sigma^2_{k})^3=\sigma^6_{k}(n_k-3)^{-3}\times E
\{\chi^2_{n_k-3}- E(\chi^2_{n_k-3})\}^3$ and 
$E(\wh\sigma^2_{k}-\sigma^2_{k})^4=\sigma^8_{k}(n_k-3)^{-4}\times E
\{\chi^2_{n_k-3}- E(\chi^2_{n_k-3})\}^4$.
For a $\chi^2$ random variable with degrees of freedom  $\scripty{r}$, 
the $s$th order raw moment is 
$\mu^{'}_s= 2^s\Gamma{(\scripty{r}/2+s)}/\Gamma{(\scripty{r}/2)}$. 
So,
$\mu^{'}_1=\scripty{r}$, $\mu^{'}_2= 4(\scripty{r}/2+1)\scripty{r}/2$,
$\mu^{'}_3= 8 (\scripty{r}/2+2)(\scripty{r}/2+1)\scripty{r}/2$, 
$\mu^{'}_4= 16 (\scripty{r}/2+3)(\scripty{r}/2+2)(\scripty{r}/2+1)\scripty{r}/2$, and these can be used to compute 
$E\{\chi^2- E(\chi^2)\}^3= \mu^{'}_3-3\mu^{'}_2\mu^{'}_1+2
(\mu^{'}_1)^3$ and 
$E\{\chi^2- E(\chi^2)\}^4= \mu^{'}_4-4\mu^{'}_3\mu^{'}_1
+6\mu^{'}_2(\mu^{'}_1)^2
-4\mu^{'}_1(\mu^{'}_1)^3
+ (\mu^{'}_1)^4
= \mu^{'}_4-4\mu^{'}_3\mu^{'}_1
+6\mu^{'}_2(\mu^{'}_1)^2
-3(\mu^{'}_1)^4
$. 
Moreover, in the final expression of $\mu_1$ and $\mu_2$, we replace $\sigma^2_k$ by its estimator $\wh\sigma^2_k$. 

In rare cases, $df$ or $c$ could be negative when computed using the formula of Theorem \ref{th1}. In that case, we propose to solve $df$ and $c$ by minimizing 
$$
\arg\min_{c>0, df>4}\left(\frac{c\mu_1}{\rho}- \frac{df}{df-2}\right)^2
+\left\{ \frac{c^2\mu_2}{\rho^2} -\frac{2df^2}{\rho(df-2)^2(df-4)}\right\}^2.
$$

When we conduct such a test for a large number of genes and each test is carried out at $\alpha$ level of significance, the familywise error rate could be much higher than $\alpha$, consequently the false discovery rate (FDR) could be high. Therefore, after calculating the p-value of the test for every gene in a set of a large number of genes,  one may use the  
Benjamini-Hochberg method \citep{Benjamini1995} or the qvalue method \citep{Storey2003} to obtain  adjusted p-values.

\subsubsection{Test for differential mesor (CARhy-TDM)}
For this test, we set 
$H_0: \delta_1=\cdots=\delta_{\mK}$ against $H_a:$ at least one of $\delta_k$ is different from the rest. 
We use the same form of test as $T_{DR}$ with $L$ being replaced by 
\begin{eqnarray*}
L_{DM}=\left(
\begin{array}{rrr rrr rrr rrr}
1 & 0 & 0 &   & -1 & 0 & 0 &   & \cdots & 0 & 0 & 0 \\
\vdots \\
1 & 0 & 0 &   & 0 & 0 & 0 &   & \cdots & -1 & 0 & 0 \\
\end{array}
\right)    
\end{eqnarray*}
and $\rho$ by $\mK-1$ and call this test statistic as $T_{DM}$.  The reference distribution of $T_{DM}$ can be obtained 
from Theorem \ref{th1}
after necessary changes of $L$ and $\rho$.

\subsubsection{Test for rhythmicity (CARhy-TR)}\label{sec:testrhythmicity}
Suppose that we are interested in the presence of rhythmicity of a gene's expression under a given condition, $k$, which can be tested by setting $H_0:\alpha^2_k+\beta^2_k=0$ versus $H_0:\alpha^2_k+\beta^2_k>0$. 
The test statistic is 
$T_R=\wh\alpha^2_k+\wh\beta^2_k$. 
Note that $(\wh \alpha_k, \wh\beta_k)^\top$ follows ${\rm normal}(0, \Lambda_k)$, where $\Lambda_k = A\sigma^2_k(X_k^\top X_k)^{-1}A^\top$, where 
\[A=\left(\begin{array}{ccc}
0 & 0 & 0 \\
0 & 1 & 0\\
0 & 0 & 1\\
\end{array}
\right).\]
Let $P_k$ be an orthonormal matrix of order $2$, then according to the spectral decomposition $\Lambda_k=P_k{\rm Diag}(\tau_{k, 1}, \tau_{k, 2})P^\top_k$.
Then 
$\wh\alpha^2_k+\wh\beta^2_k= (\wh\alpha_k, \wh\beta_k)^\top P_k P^\top_k (\wh\alpha_k, \wh\beta_k)= \{P^\top_k(\wh\alpha_k, \wh\beta_k)\}^\top P^\top_k(\wh\alpha_k, \wh\beta_k)$. Observe that, under $H_0$, $(U_{k, 1}, U_{k, 2})^\top=P^\top_k(\wh\alpha_k, \wh\beta_k)$ follows ${\rm Normal}(0, {\rm Diag}(\tau_{k, 1}, \tau_{k, 2}))$. So,  
$\wh\alpha^2_k+\wh\beta^2_k=U^2_{k, 1}+U^2_{k, 2}$, where 
$U^2_{k, 1}/\tau_{k, 1}$ and $U^2_{k, 2}/\tau_{k, 2}$ follow independent $\chi^2_1$ distribution ($\chi^2$ distribution with degrees of freedom one). So, under $H_0$, $\wh\alpha^2_k+\wh\beta^2_k$ follows a weighted sum of two independent $\chi^2_1$ distributions with weights 
$\tau_{k, 1}$ and $\tau_{k, 2}$. So, $T_R$ is equivalent to 
\[
 \chi^2_1\tau_{k, 1}+\chi^{*, 2}_1\tau_{k, 2}
\] in distribution,  where $\chi^2_1$ and $\chi^{*, 2}_1$ are two independent $\chi^2_1$ random variables.
However, enumeration of the above is difficult as $\tau_{k, 1}$ and $\tau_{k, 2}$ are unknown. Since $(n_k-3)\wh\tau_{k, 1}\sim \tau_{k, 1}\chi^2_{n_k-3}$
and $(n_k-3)\wh\tau_{k, 2}\sim \tau_{k, 2}\chi^2_{n_k-3}$, after replacing 
$\tau_{k, 1}$ and 
$\tau_{k, 2}$
by $(n_k-3)\wh\tau_{k, 1}/\chi^2_{n_k-3}$
and 
$(n_k-3)\wh\tau_{k, 2}/\chi^2_{n_k-3}$, 
under $H_0$,
\[
\wh\alpha^2_k+\wh\beta^2_k\stackrel{d}{=} \chi^2_1\tau_{k, 1}+\chi^{*, 2}_1\tau_{k, 2}{ \approx} \frac{\chi^2_1(n_k-3)\wh\tau_{k, 1}}{\chi^2_{n_k-3}}+
\frac{\chi^{*,2}_1(n_k-3)\wh\tau_{k, 2}}{\chi^2_{n_k-3}}.
\]
\paragraph{Monte Carlo p-value computation.}
Generate $U_{i}\sim\chi^2_1$ for $i=1,\ldots,2N$ and
$U^*_{i}\sim\chi^2_{n_k-3}$ for $i=1,\ldots,N$.
For $i=1,\ldots,N$, compute
\[
V_i=\frac{(n_k-3)\big(U_{i}\widehat{\tau}_{k,1}
+U_{i+N}\widehat{\tau}_{k,2}\big)}{U^*_{i}}.
\]
The p-value is then estimated by  
$N^{-1}\sum_{i=1}^N
I\!\left(V_i>
\widehat{\alpha}_k^2+\widehat{\beta}_k^2\right)$, where $I$ denotes the indicator function.  
 Although we did not find any indication of inflated type-I error rate in our simulation studies, the estimated $p$-value from this Monte Carlo procedure is likely to be slightly conservative, in the sense that it tends to be larger than the true $p$-value. This occurs because the simulated test statistic incorporates additional sources of variability beyond those present in $T_R = \widehat{\alpha}_k^2 + \widehat{\beta}_k^2$, resulting in a null distribution with inflated variance. Consequently, the Monte Carlo-generated distribution is likely to place more mass in the tails, making the observed value of $T_R$ appear less extreme than it would under its true sampling distribution.

\subsubsection{Tests for differential amplitude and differential phase}\label{DAnDP}
If a gene exhibits differential rhythmicity across conditions, we may further investigate whether this difference is attributable to differential amplitude (DA) or differential phase (DP).

\subsubsection*{Differential Amplitude (CARhy-TDA)}

To test for differential amplitude, we consider the hypotheses
\[
H_0:\ \theta_1=\cdots=\theta_{\mK}
\quad \text{versus} \quad
H_a:\ \text{at least one } \theta_k \text{ differs},
\]
where $\theta=(\theta_1,\ldots,\theta_{\mK})^\top$ and
$\theta_k = (\alpha_k^2+\beta_k^2)^{1/3}$. 
We adopt the cube-root transformation instead of the conventional amplitude
$(\alpha_k^2+\beta_k^2)^{1/2}$ because the estimator
$(\wh\alpha_k^2+\wh\beta_k^2)^{1/3}$
exhibits improved convergence to a normal distribution
\citep{Wilson1931}.
The estimator of $\theta_k$ is
$
\wh\theta_k = (\wh\alpha_k^2+\wh\beta_k^2)^{1/3}
$,  
and an approximate variance of $\wh\theta_k$ is 
$
D_k^\top \wh\Sigma_k D_k$, 
where
\[
D_k =
\Bigl(
0,\;
\tfrac{2}{3}\,\alpha_k(\alpha_k^2+\beta_k^2)^{-2/3},\;
\tfrac{2}{3}\,\beta_k(\alpha_k^2+\beta_k^2)^{-2/3}
\Bigr)^\top.
\]
In practice, $D_k$ is replaced by its plug-in estimator obtained by substituting
$(\alpha_k,\beta_k)$ with $(\wh\alpha_k,\wh\beta_k)$. 
The differential amplitude (DA) test statistic is
\[
T_{DA}
=
\frac{1}{\rho}
(L\wh\theta)^\top
\Bigl\{
L\,\mathrm{Diag}
\bigl(
D_1^\top\wh\Sigma_1D_1,\ldots,
D_\mK^\top\wh\Sigma_\mK D_\mK
\bigr)
L^\top
\Bigr\}^{-1}
(L\wh\theta),
\]
where $\wh\theta=(\wh\theta_1, \ldots, \wh\theta_{\mK})^\top$ and  $L$ is a $(\mK-1)\times\mK$ contrast matrix defined as
\begin{equation}\label{eq:L2}
L=
\begin{pmatrix}
1 & -1 & 0 & \cdots & 0 \\
1 & 0  & -1 & \cdots & 0 \\
\vdots & & & \ddots & \\
1 & 0  & 0 & \cdots & -1
\end{pmatrix},
\end{equation}
and $\rho=\mathrm{rank}(L)=\mK-1$.

\subsubsection*{Differential Phase (CARhy-TDP)}

To test for differential phase, we consider the same hypotheses,
$
H_0: \theta_1=\cdots=\theta_{\mK}$
versus $H_a: \mbox{at least one } \theta_k \mbox{ differs},
$
but define
$
\theta_k = \mathrm{atan2}(\beta_k,\alpha_k)$. 
The DP test statistic, denoted by $T_{DP}$, is obtained from $T_{DA}$ after replacing
$
\wh\theta_k$ by $\mathrm{atan2}(\wh\beta_k,\wh\alpha_k)
$
and
\[
D_k =
\Bigl(
0,\;
-\wh\beta_k(\wh\alpha_k^2+\wh\beta_k^2)^{-1},\;
\wh\alpha_k(\wh\alpha_k^2+\wh\beta_k^2)^{-1}
\Bigr)^\top.
\]
The contrast matrix $L$ remains unchanged.

\subsubsection*{Null Distribution}
Importantly, in both settings, the inferential problem is substantially more complex than a routine plug-in procedure. Without invoking large-sample theory or homoskedastic error structures, the sampling distribution of these statistics must be derived with care, accounting for condition-specific variability and the nonlinear dependence of the parameters on the underlying regression coefficients. The null distributions of $T_{DA}$ and $T_{DP}$ are given in the following result.

\begin{Th}\label{th2}
Under the stated assumptions and when $\alpha_k^2+\beta_k^2>0$ for all $k$, the distribution of $c\,T_{DA}$ or $c\,T_{DP}$ is approximately an $F$ distribution with numerator degrees of freedom $\rho$ and denominator degrees of freedom $df$, where
\[
df = \frac{\rho-2 + 2\rho\mu_2/\mu_1^2}{\rho\mu_2/(2\mu_1^2)-1},
\qquad
c = \frac{\rho\,df}{\mu_1(df-2)}.
\]
Here,
\[
\mu_1 = \rho + \sum_{k=1}^{\mK} \frac{2\wh\sigma_k^4}{n_k-3}\,\mathrm{tr}(G_kG_k),
\]
and
\begin{align*}
\mu_2
&= 2\rho
+ \sum_{k=1}^{\mK} \var(\wh\sigma_k^2)
\Bigl[
\{\mathrm{tr}(G_k)\}^2 + 6\,\mathrm{tr}(G_kG_k)
\Bigr] \\
&\quad
- \sum_{k=1}^{\mK}
E\bigl\{(\wh\sigma_k^2-\sigma_k^2)^3\bigr\}
\Bigl[
2\,\mathrm{tr}(G_k)\mathrm{tr}(G_kG_k)
+4\,\mathrm{tr}(G_kG_kG_k)
\Bigr] \\
&\quad
+ \sum_{k=1}^{\mK}
E\bigl\{(\wh\sigma_k^2-\sigma_k^2)^4\bigr\}
\Bigl[
\{\mathrm{tr}(G_kG_k)\}^2
+2\,\mathrm{tr}(G_kG_kG_kG_k)
\Bigr] \\
&\quad
- \sum_{k=1}^{\mK} \var^2(\wh\sigma_k^2)\{\mathrm{tr}(G_kG_k)\}^2 \\
&\quad
+ \sum_{s>k}
\var(\wh\sigma_k^2)\var(\wh\sigma_s^2)
\Bigl[
\{\mathrm{tr}(G_kG_s+G_sG_k)\}^2
+4\,\mathrm{tr}(G_kG_kG_sG_s) \\
&\hspace{25mm}
+2\,\mathrm{tr}\bigl\{(G_kG_s+G_sG_k)^2\bigr\}
\Bigr].
\end{align*}

The matrices $G_k$ are defined as
\[
G_k =
L\,\mathrm{Diag}(0,\ldots,D_k^\top(X_k^\top X_k)^{-1}D_k,\ldots,0)\,L^\top\,\Psi^{-1},
\]
where
\[
\Psi
=
L\,\mathrm{Diag}
\bigl(
D_1^\top\Sigma_1D_1,\ldots,
D_\mK^\top\Sigma_\mK D_\mK
\bigr)
L^\top.
\]
\end{Th}

The proof of Theorem~\ref{th2} is provided in Supplementary Appendix~\ref{app:proof-thm2}.
The vectors $D_k$ take different forms depending on whether $T_{DA}$ or $T_{DP}$ is considered.
Both tests should be conducted only when rhythmicity is present under all conditions. The presence of rhythmicity can be checked via the technique of Section \ref{sec:testrhythmicity}.

\section{Simulation Studies}
\subsection{Simulation design}
We assessed and compared our methods against two competing methods: DODR, based on hierarchical ANOVA, and dryR based on a Bayesian framework. 
First  we focus was on the test of differential rhythmicity (TDR).

To assess the type-I error rate for the test of DR, we simulated a gene's expression data under $H_0$ that there is no differentially rhythmicity across two and three experimental conditions, separately. We repeated this procedure for $R=10,000$ times. Every time, we calculated the test statistic and estimated the p-value using the proposed method, and recorded the proportion of times  $H_0$ was rejected  out of $R$ replications. For checking power, we simulated data under $H_a$, where the gene expression followed differentially rhythmicity.  

Following our real data, we simulated expression profile from the model 
\begin{eqnarray} \label{eq:simu}
Y_{j,k}(t)=1+  A_k\cos\left( \frac{2\pi}{24}(t -\phi_k)\right)+
\epsilon_{j, k}(t)
\end{eqnarray}
at time points $t= 0, 4, 8, 12, 16, 20$ hr, for $k=1, \dots, \mK$ and $j=1, \cdots, \mJ_{i, k}$. 
We simulated $\epsilon_{j, k}(t)\stackrel{iid}{\sim} {\rm Normal}(0, \sigma^2_k)$. We considered scenarios 1) where $\sigma^2_k$ was varying over conditions and 2) where $\sigma^2_k$ was fixed over different conditions. The parameter values for simulation are given in 
Tables \ref{table:table1} and \ref{table:table2}. Additionally, we considered unbalanced sampling scenarios, as shown in Figures \ref{fig:figure1} and \ref{fig:figure2}. 

To assess the robustness of our method to heteroskedastic noise, we simulated data where $\epsilon$ follows homoscedastic Gaussian distribution (Cases 1, 5 of Table \ref{table:table1} and Case 9, 13 of Table \ref{table:table2}) or a heteroskedastic Gaussian distribution (Cases 2, 6 of Table \ref{table:table1} and Case 10, 14 of Table \ref{table:table2}). To assess the sensitivity of our method to the normality assumption of noise, we simulated data where $\epsilon$ follows a Student's t-distribution with 3 or 4 degrees of freedom (Cases 3, 4, 7, 8 of Table \ref{table:table1} and Case 11, 12, 15, 16 of Table \ref{table:table2}).

\begin{table}[!htbp]
\centering
\renewcommand{\arraystretch}{1.0}
\caption{Parameter values for non-differential rhythmicity scenarios in the simulation.}
\begin{tabular}{cc ccc ccc ccc ccc}
\toprule
\multirow{2}{*}{Group} & \multirow{2}{*}{Case}
& \multicolumn{3}{c}{Amplitude}
& \multicolumn{3}{c}{Phase}
& \multicolumn{3}{c}{Normal dist.}
& \multicolumn{3}{c}{Student's $t$ dist.} \\
\cmidrule(lr){3-5}\cmidrule(lr){6-8}\cmidrule(lr){9-11}\cmidrule(lr){12-14}
& & $A_1$ & $A_2$ & $A_3$ & $\phi_1$ & $\phi_2$ & $\phi_3$
& $\sigma_1$ & $\sigma_2$ & $\sigma_3$ & $df_1$ & $df_2$ & $df_3$ \\
\midrule

\multirow{4}{*}{3-conditions}
& 1 & 1 & 1 & 1 & 5 & 5 & 5 & 1   & 1 & 1 & -- & -- & -- \\
& 2 & 1 & 1 & 1 & 5 & 5 & 5 & 0.5 & 2 & 2 & -- & -- & -- \\
& 3 & 1 & 1 & 1 & 5 & 5 & 5 & 1   & --& --& -- & $0.4t_3$ & $0.4t_3$ \\
& 4 & 1 & 1 & 1 & 5 & 5 & 5 & --  & --& --& $t_4$ & $0.1t_4$ & $0.5t_4$ \\
\midrule

\multirow{4}{*}{2-conditions}
& 5 & 1 & 1 & -- & 5 & 5 & -- & 1   & 1 & -- & -- & -- & -- \\
& 6 & 1 & 1 & -- & 5 & 5 & -- & 0.5 & 2 & -- & -- & -- & -- \\
& 7 & 1 & 1 & -- & 5 & 5 & -- & 1   & --& -- & -- & $0.4t_3$ & -- \\
& 8 & 1 & 1 & -- & 5 & 5 & -- & --  & --& -- & $t_4$ & $0.5t_4$ & -- \\
\bottomrule
\label{table:table1}
\end{tabular}
\end{table}

\begin{table}[!htbp]
\centering
\renewcommand{\arraystretch}{1.0}
\caption{
Parameter values for differential rhythmicity scenarios in the simulation.}
\begin{tabular}{cc ccc ccc ccc ccc}
\toprule
\multirow{2}{*}{Model} & \multirow{2}{*}{Case}
& \multicolumn{3}{c}{Amplitude}
& \multicolumn{3}{c}{Phase}
& \multicolumn{3}{c}{Normal dist.}
& \multicolumn{3}{c}{Student's $t$ dist.} \\
\cmidrule(lr){3-5}\cmidrule(lr){6-8}\cmidrule(lr){9-11}\cmidrule(lr){12-14}
& & $A_1$ & $A_2$ & $A_3$
  & $\phi_1$ & $\phi_2$ & $\phi_3$
  & $\sigma_1$ & $\sigma_2$ & $\sigma_3$
  & $df_1$ & $df_2$ & $df_3$ \\
\midrule

\multirow{4}{*}{3-conditions}
& 9  & 1 & 1 & 1.5 & 5 & 2.5 & 5 & 1   & 1   & 1   & --  & --        & -- \\
& 10 & 1 & 1 & 1.5 & 5 & 2.5 & 5 & 0.5 & 2   & 2   & --  & --        & -- \\
& 11 & 1 & 1 & 1.5  & 5 & 2.5 & 5 & 1   & --  & --  & --  & $0.4t_3$   & $0.4t_3$ \\
& 12 & 1 & 1 & 1.5   & 5 & 2.5 & 5 & --  & --  & --  & $t_4$ & $0.1t_4$ & $0.5t_4$ \\
\midrule

\multirow{4}{*}{2-conditions}
& 13  & 1 & 1.5 & -- & 5 & 2.5 & -- & 1   & 1   & -- & --   & --       & -- \\
& 14 & 1 & 1.5 & -- & 5 & 2.5 & -- & 0.5 & 2   & -- & --   & --       & -- \\
& 15 & 1 & 1.5   & -- & 5 & 2.5  & -- & 1   & --  & -- & --   & $0.4t_3$  & -- \\
& 16 & 1 & 1.5   & -- & 5 & 2.5  & -- & --  & --  & -- & $t_4$ & $0.5t_4$ & -- \\
\bottomrule
\label{table:table2}
\end{tabular}
\end{table}


\subsection{Type I error rate and power analysis}

DODR tests differential rhythmicity between two groups by fitting harmonic regressions and applying nested ANOVA to compare the model with shared rhythm to a model with both shared rhythm and differential rhythm. dryR performs model selection over a candidate set using Bayesian Information Criterion weights (BICW),  and the model with highest BIC weight is selected. It classifies a gene as differentially rhythmic if the selected model is outside the null model set and its BIC weight is greater than a specific threshold (0.95 for 2 conditions as reported by Weger et al.). We employed three thresholds of 0.7, 0.8, and 0.9 in this simulation study, which were respectively named dryR\_0.7, dryR\_0.8 and dryR\_0.9.

\begin{figure}[!htbp]
\centering
\includegraphics[height=10cm, width=15cm]{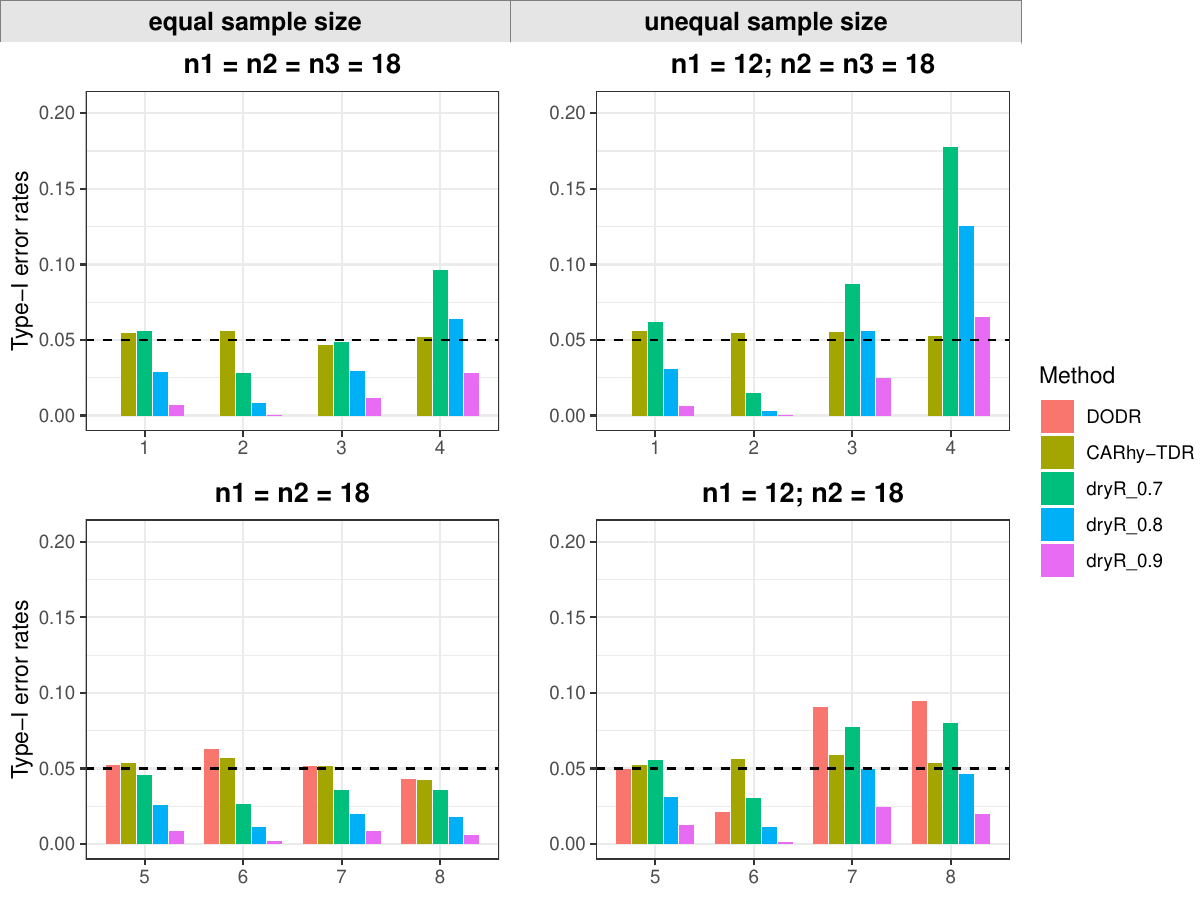}
\caption{Type-I error rates for Cases 1-8 of Table \ref{table:table1}.}
\label{fig:figure1}
\end{figure}

Figure~\ref{fig:figure1} presents the type I error rates for the cases (simulation scenarios) of  Table~\ref{table:table1}. CARhy-TDR maintains the type I error rate close to the nominal level of 0.05 across all cases with both homoscedastic and heteroscedastic noise, both equal and unequal sample size scenarios (i.e., unbalanced design), and both Gaussian and heavy-tailed noise, indicating strong robustness to heteroscedastic noise, heavy-tailed noise, and unbalanced sampling.
By contrast, under two conditions, DODR shows a significantly inflated Type I error rate in the unequal sample size scenario with heavy-tailed noise; the same pattern is also observed for dry.R when a BICW threshold of 0.7 is used (Case 7 and Case 8). Under three conditions, BICW thresholds of 0.7 and 0.8 provide acceptable control of the Type I error rate under homoscedasticity, but show substantial inflation of the Type I error rate under heteroscedasticity and unequal sampling (Case 3 and Case 4). Choosing a BICW threshold of 0.9 keeps the Type I error rate below 0.05 in most cases; however, under heteroscedasticity with unbalanced sampling, the Type I error rate is still inflated (Case 4).

\begin{figure}[!htbp]
\centering
\includegraphics[height=10cm, width=15cm]
{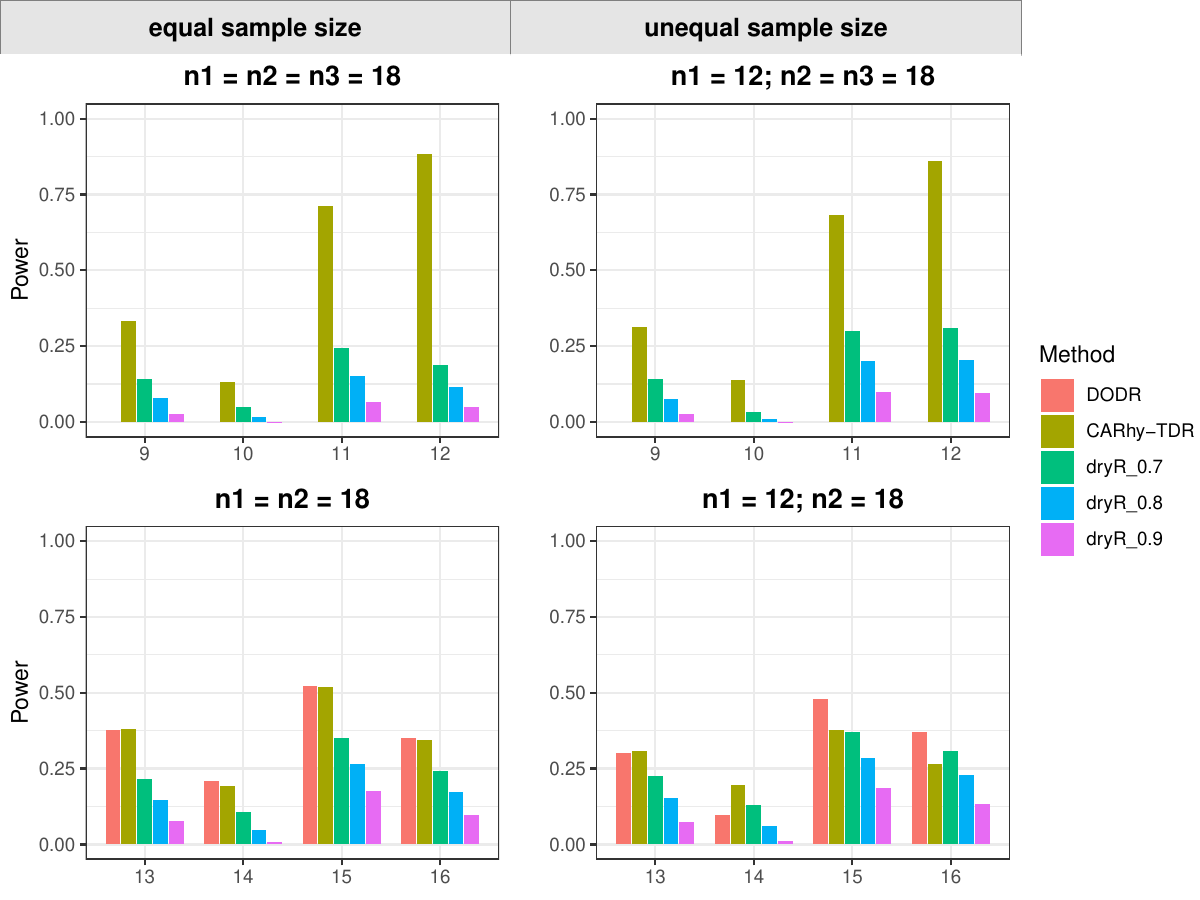}
\caption{Power for Cases 9-16 of Table \ref{table:table2}.}
\label{fig:figure2}
\end{figure}

Figure~\ref{fig:figure2} presents the powers for the cases of  Table~\ref{table:table2}. Under three experimental conditions, CARhy-TDR achieves the highest power in all cases, outperforming dryR at any prespecified BICW threshold, with a particularly advantage under heavy-tailed noise (Case 11 and Case 12); meanwhile, its type I error rate remains close to the nominal 0.05 level (Case 3 and Case 4). Under two experimental conditions with equal sample size, CARhy-TDR and DODR have the highest and comparable power; however, under unequal sample sizes with heteroscedastic noise (Case 14), CARhy-TDR performs more powerful. Under unequal sample size with heavy-tailed noise, DODR shows higher power but at the cost of a markedly inflated type I error rate above the nominal 0.05 level (Case 7 and Case 8).

\subsection{FDR control and F1 score analysis for the differential rhythmicity test}

We conducted simulations with 2,000 genes, of which 10\% were designated as truly differentially rhythmic (DR) and the remaining genes as non-DR. We evaluated performance under both two experimental conditions (group 1 vs. group 2) and three experimental conditions (group 1 vs. group 2 vs. group 3), using the parameter settings as in Table~\ref{table:table3}.

\begin{table}[!htbp]
\centering
\renewcommand{\arraystretch}{1.0}
\caption{
Parameter values for the 
simulation scenarios for FDR and F1 evaluations.}
\begin{tabular}{@{}p{4cm}p{3.5cm}p{7cm}@{}}
\toprule
\textbf{Parameter} & \textbf{Value} & \textbf{Description} \\
\midrule
$sd_1$ & 0.5 & noise standard deviation for group 1 \\
$sd_2=sd_3$ & 1.0 & noise standard deviation for group 2, 3 \\
$A_{\mathrm{DR (1)}}$ & 1.0 & differential amplitude for group 1 \\
$A_{\mathrm{DR (2)}}=A_{\mathrm{DR (3)}}$ & 2.0 & differential amplitude for group 2, 3 \\
$\phi_{\mathrm{DR (1)}}$  & 5.0 & differential phase for group 1 \\
$\phi_{\mathrm{DR (2)}}=\phi_{\mathrm{DR (3)}}$  & 10.0 & differential phase for group 2, 3 \\
$A_{\mathrm{nonDR}}$  & 1.0 & shared amplitude for non-DR\\
$\phi_{\mathrm{nonDR}}$  & 5.0 & shared phase for non-DR\\
$\mJ_{i, 1}$  & 2 & replicates per time point for group 1\\
$\mJ_{i, 2}= \mJ_{i, 3}$  & 3 & replicates per time point for group 2, 3\\
Sampling time points & 0, 4, 8, 12, 16, 20 & measurement time points\\
\bottomrule
\label{table:table3}
\end{tabular}
\end{table}


For each condition, gene expression was generated from  model (\ref{eq:simu}) with heteroscedastic Gaussian noise with a mean of zero. Measurements were taken at six time points and replication at each time point was unbalanced across conditions. Non-DR genes shared identical rhythmic parameters (amplitude and phase) across all conditions, whereas DR genes used condition-specific rhythmic parameters across all conditions. We applied CARhy-TDR and DODR to the data and p values were adjusted by Benjamini-Hochberg method. We also applied dryR with three thresholds, 0.7 (dryR\_0.7), 0.8 (dryR\_0.8), and 0.9 (dryR\_0.9). 
We repeated this simulation 30 times, and 
each simulated data contained  2000 genes. For each simulated data we calculated the false discovery rate (FDR) and F1 score: 

\[
\mathrm{FDR}=\frac{\mathrm{FP}}{\mathrm{TP}+\mathrm{FP}},\qquad
\mathrm{F1}=\frac{2\,\mathrm{TP}}{2\,\mathrm{TP}+\mathrm{FP}+\mathrm{FN}},
\]
 under each method, where 
FP, FN, TP denote the number of false positives, the number of false negatives, and the number of true positives. 

\begin{figure}[!htbp]
\centering
\includegraphics[width=0.88\textwidth]{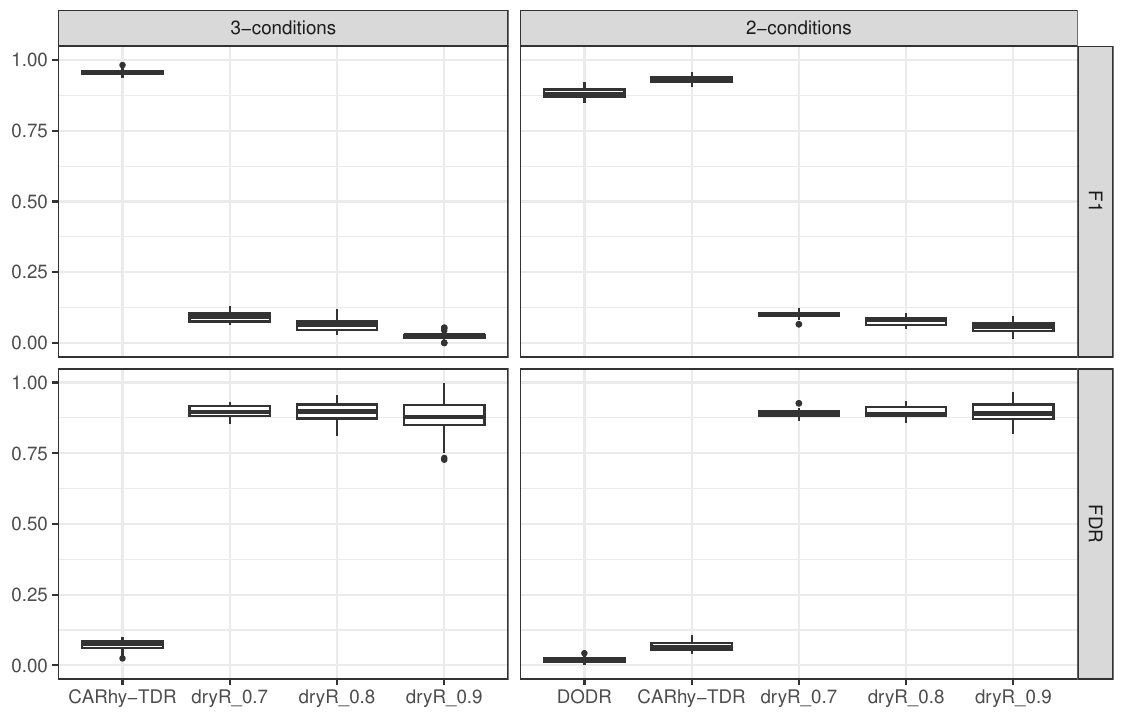}
\caption {F1 scores (upper panel) and FDR (lower panel)  for different methods. For DODR and CARhy-TDR, we adjusted the p-values using the Benjamini-Hochberg method. 
}  
\label{fig:figure3}
\end{figure}

Figure~\ref{fig:figure3} shows the performance of the methods under  these metrics.  CARhy-TDR exhibited the best performance compared to dryR under all preset BICW thresholds, achieving the highest F1 score while maintaining the lowest FDR, indicating a robust balance between sensitivity and false positive control under the scenario where genes are independent.  A similar trend was observed under the design of two experimental conditions. CARhy-TDR achieved a higher F1 score than DODR while keeping the FDR around 0.05, indicating a favorable balance between detection performance and false discovery control.  The results highlight opportunities for improvement in the dryR method, suggesting that its decision rule (e.g., the BIC-weight threshold) could benefit from further calibration to enhance performance.

\subsection{Performance of other tests }
\subsubsection{Design} 
 To evaluate the performance of the test of rhythmicity (CARhy-TR) and compare with the nonparametric test JTK\_CYCLE \citep{PMID20876817} for a single gene under one condition. Data were generated according to
$
Y_{j}(t_i) = 1 + A \cos\!\left\{ \frac{2\pi}{24}(t_i - \phi) \right\} + \epsilon_{i, j},
$
with varying signal strengths, $A=0, 0.5, 1, 2$, $\phi=5$,  $\epsilon_{i, j}\stackrel{iid}{\sim} {\rm Normal}(0, 1)$ and $\epsilon_{i, j}\stackrel{iid}{\sim} t_3$ (for heavy tail) for two sample sizes $12$ ($n_k=6\times 2 = 12$) and $18$ ($n_k=6\times 3 = 18$).

To assess type I error rates and power for testing differential mesor (CARhy-TDM), we conducted simulation studies under two- and three-condition experimental designs. For each setting, 10{,}000 datasets were generated using the model
$
Y_{j,k}(t_i) = \delta_k + A \cos\!\left\{ \frac{2\pi}{24}(t_i - \phi) \right\} + \epsilon_{j,k}(t_i),
$ 
where $k$ indexes experimental conditions, $i$ denotes sampling times (in hours) over a 24-hour cycle, and $j$ indexes replicates. Parameter values are provided in Table~\ref{tab:S1}, and simulations were performed for both sample size settings ($n_k = 12$ and $n_k = 18$). Empirical rejection rates across simulations were used to estimate type I error under the null hypothesis and power under the alternative.

For testing differential amplitude (CARhy-TDA), simulations were conducted using parameter settings in Table~\ref{tab:S2}, with gene expression generated from
$
Y_{j,k}(t_i) = 1 + A_k \cos\!\left\{ \frac{2\pi}{24}(t_i - \phi) \right\} + \epsilon_{i, j,k}.
$
Finally, to assess performance for testing differential phase (CARhy-TDP), simulations were conducted using parameter settings in Table~\ref{tab:S3}. Gene expression data were generated as
$
Y_{j,k}(t_i) = 1 + A \cos\!\left\{ \frac{2\pi}{24}(t_i - \phi_k) \right\} + \epsilon_{i, j,k}.
$

\subsubsection{Results and discussion}

\begin{figure}[!htbp]
\centering
\includegraphics[angle=90,height=0.98\textheight,keepaspectratio]
{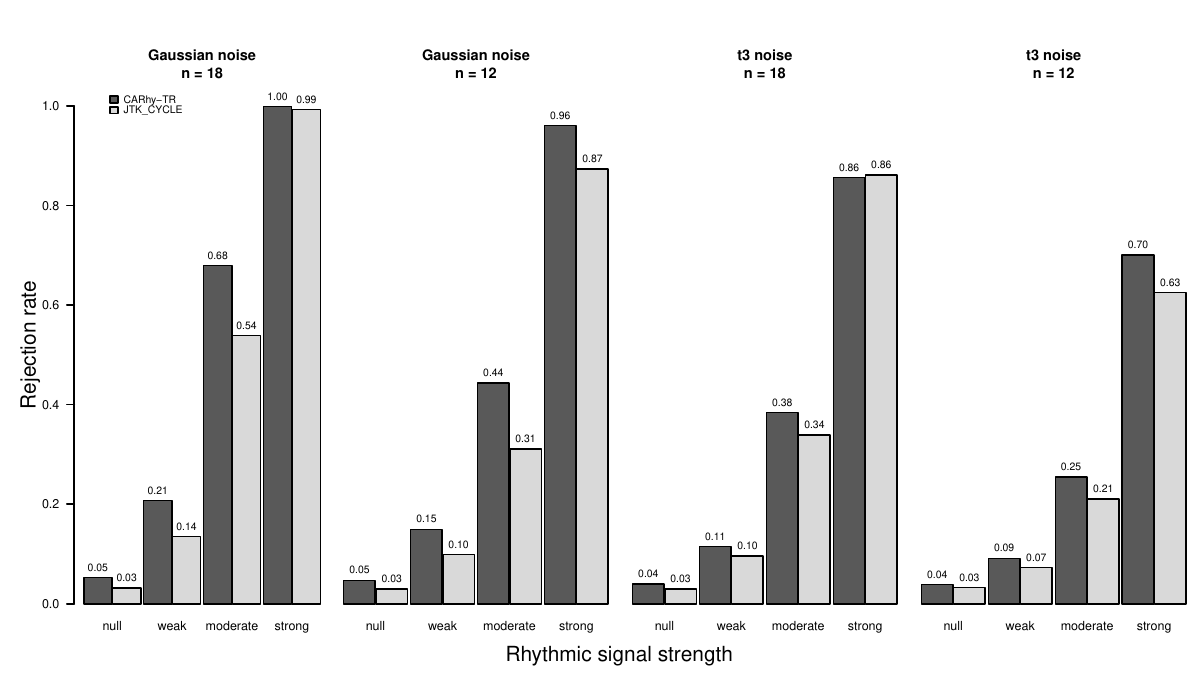}
\caption{Type-I error rate and power for testing rhythmicity. Black: CARhy-TR, Gray: JTK\_CYCLE}
\label{fig:S1}
\end{figure}

Figure~\ref{fig:S1} illustrates the performance of the proposed test for detecting rhythmicity (CARhy-TR). Under the null setting, CARhy-TR maintained the type I error rate close to the nominal level of 0.05, whereas JTK\_CYCLE was more conservative. For all the weak, moderate and strong rhythmic signals, CARhy-TR showed higher detection power than JTK\_CYCLE with both sample sizes 12 and 18; and it even showed higher power when testing signals with heavy-tail noises, especially weak and moderate signals.

Figure~\ref{fig:S2} shows the performance of testing differential mesor (CARhy-TDM). In both two- and three-condition designs, CARhy-TDM maintained the type I error close to the nominal level of 0.05, under both homoskedastic and heteroskedastic noise scenarios; meanwhile CARhy-TDM exhibits high power for all scenarios with both sample sizes 12 and 18.

Figure~\ref{fig:S3} summarizes the performance of the proposed test for differential amplitude (CARhy-TDA). We compared CARhy-TDA with DODR, dryR, and CARhy-TDR that assess differential rhythmicity without distinguishing between amplitude and phase effects. For dryR, we used three BICW thresholds (0.7, 0.8, and 0.9), denoted as dryR\_0.7, dryR\_0.8, and dryR\_0.9, respectively. In the three-condition scenarios, CARhy-TDA maintained the type I error rate close to the nominal 0.05 level and was slightly lower than that of CARhy-TDR. In terms of power, CARhy-TDA was slightly more powerful than CARhy-TDR and consistently outperformed dryR at all preset BICW thresholds. This advantage was particularly pronounced under heteroscedasticity (Case 4 and Case 8). Compared with DODR, CARhy-TDA exhibited a comparable type I error rate but achieved higher power across all settings, with the largest gains observed under heteroskedastic noise and unequal sample sizes (Case 8).

Figure~\ref{fig:S4} shows the performance of testing differential phase (CARhy-TDP). Across the three-condition scenarios, CARhy-TDP maintained the type I error rate well and had a slightly lower type I error rate than CARhy-TDR. In terms of power, CARhy-TDP was slightly more powerful than CARhy-TDR and consistently outperformed dryR at all preset BICW thresholds. This advantage was particularly pronounced under heteroscedasticity (Case 12 and Case 16). Compared to DODR, CARhy-TDP exhibited similar type I error control but consistently higher power across all scenarios, with the advantage being most evident under heteroskedastic and unbalanced setting (Case 16).

Overall, these results indicate that CARhy provides well-controlled type I error and superior power across a range of scenarios, especially in more complex or realistic settings, making it a robust and reliable choice for detecting differential rhythmicity, particularly differences in rhythmic components such as amplitude and phase.

\section{Application and Assessment on Transcriptomic Data}\label{section:4}
We have introduced the dataset in Section~\ref{sec2:maa}, which consists of three experimental conditions ($\mathcal{K}=3$), common sampling times $t_{1,k}=2$, $t_{2,k}=6$, $t_{3,k}=10$, $t_{4,k}=14$, $t_{5,k}=18$, and $t_{6,k}=22$, and a balanced design with $\mathcal{J}_{i,k}=3$ biological replicates at each time point for $i=1,\ldots,6$ and $k=1,2,3$, and  objectives. 

In the dataset, sequenced reads were pre-processed, aligned to the mouse genome, and gene expression was quantified using transcripts per million (TPM) normalization. Normalization to gene length was not performed because reads originate from 3'-mRNA sequencing. Genes with TPM $>$ 1 in at least half of the samples were retained, and TPM values were transformed into log2(TPM + 1) for all analyses.

\begin{figure}[!htbp]
\centering
\includegraphics[angle=90,height=0.98\textheight,keepaspectratio]{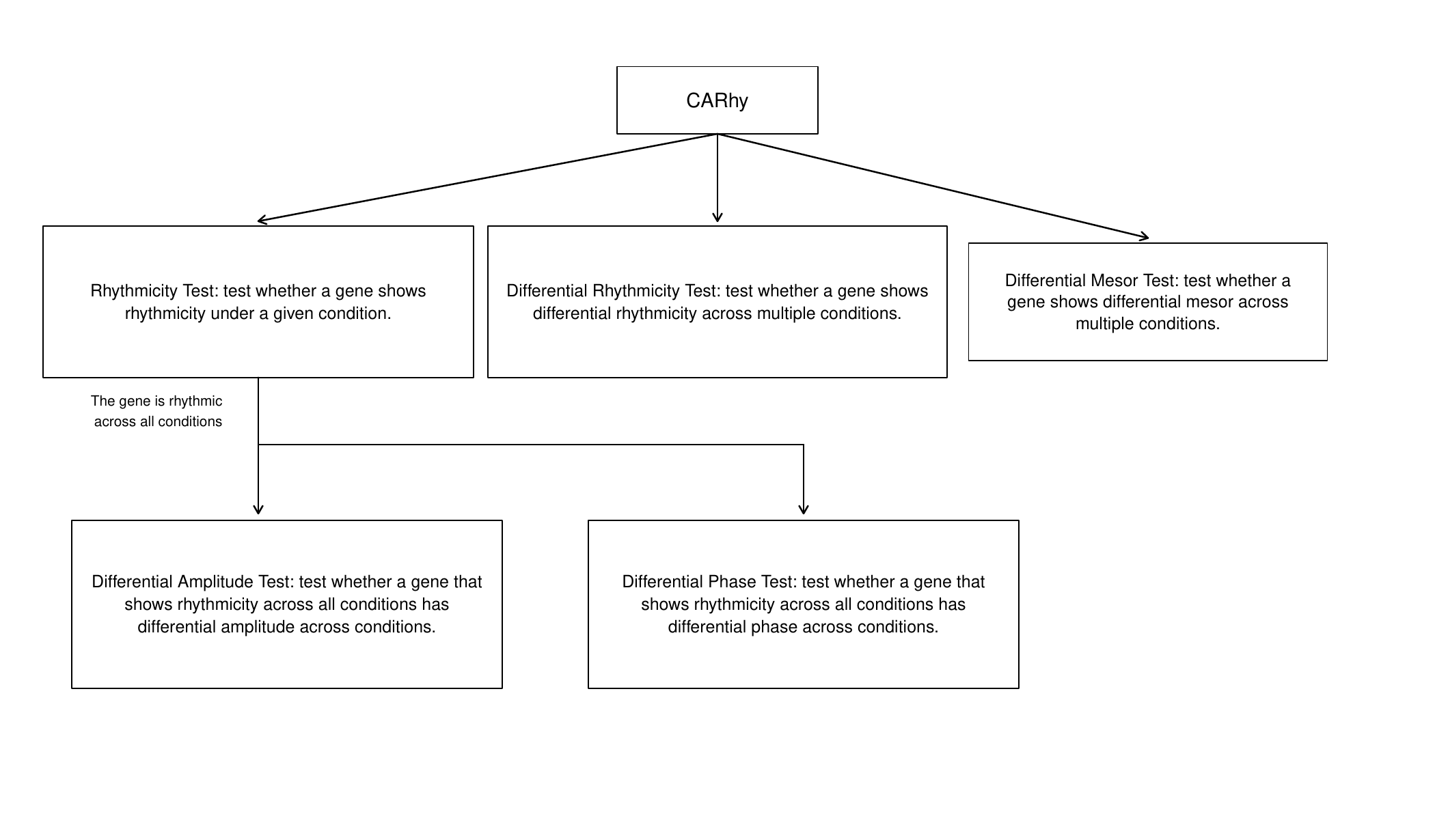}
\caption{CARhy's logic diagram.}
\label{fig:workflow}
\end{figure}

After pre-processing, we are left with 12,365 genes shared across all conditions. We analyzed these genes using CARhy. Figure~\ref{fig:workflow} illustrates the logic diagram of CARhy, including its main functions for testing rhythmicity (CARhy-TR) and differential rhythmicity (CARhy-TDR), as well as modules for testing differential amplitude (CARhy-TDA), differential phase (CARhy-TDP), and other related features. Detailed statistical results are summarized in the Table S4, including gene-specific raw p-values for all the statistical tests in CARhy, along with mesor, amplitude, and phase estimates for each condition.

Among these 12,365 genes, CARhy-TR identified 4,606 (NF), 3,870 (LB), and 3,242 (AF) rhythmic genes. Of these 12,365 genes, 1,193 genes showing rhythmicity under all three conditions with a p-value $<0.05$. Among the 1,193 genes, CARhy-TDR identified 469 differential rhythmic (DR) genes with a p-value $<0.05$ and 724 non-differential rhythmic (NDR) with a p-value $\geq$ 0.05. Figure~\ref{fig:S8} illustrates the standardized temporal expression profiles of these DR and NDR genes. NDR genes exhibited relatively similar profiles across conditions, whereas DR genes exhibited very mild rhythmicity in AF.

For validating the proposed  method for the real data analysis, we   compare between-condition differences of rhythmic features for DR genes and NDR genes. To assess the consistency of temporal expression patterns across conditions, we converted gene expression profiles to z-score for each gene within each condition across the six time points. For each gene, Pearson correlation coefficients were then calculated for all three pairwise condition comparisons (NF-LB, NF-AF, and LB-AF) and the mean of the three correlation coefficients was used as the overall temporal expression similarity metric for each gene. To quantify the amplitude divergence, absolute differences in the estimated amplitude were calculated for the three pairwise condition comparisons, and their mean was used as the overall amplitude difference metric for each gene. To quantify phase dispersion, pairwise phase differences were calculated using circular distance with a 24-hour period (the minimal circular distance), and the mean of the three pairwise differences was used as the overall phase difference metric for each gene. 
For the amplitude and phase dispersion, we  use the model-based estimates in terms of $\alpha$ and $\beta$ parameters.  
The distributions of these three metrics were compared between DR and NDR genes using two-sided Wilcoxon rank-sum tests. Finally, to compare average expression trajectories, the three biological replicates were first averaged at each time point within each condition. The resulting 18 values (6 time points $\times$ 3 conditions) for each gene were then standardized using z-scores, and the mean $\pm$ standard error of the mean was calculated across genes at each time point separately for the DR and NDR groups.

\begin{figure}[!htbp]
\centering
\includegraphics[angle=90,height=0.98\textheight,keepaspectratio]{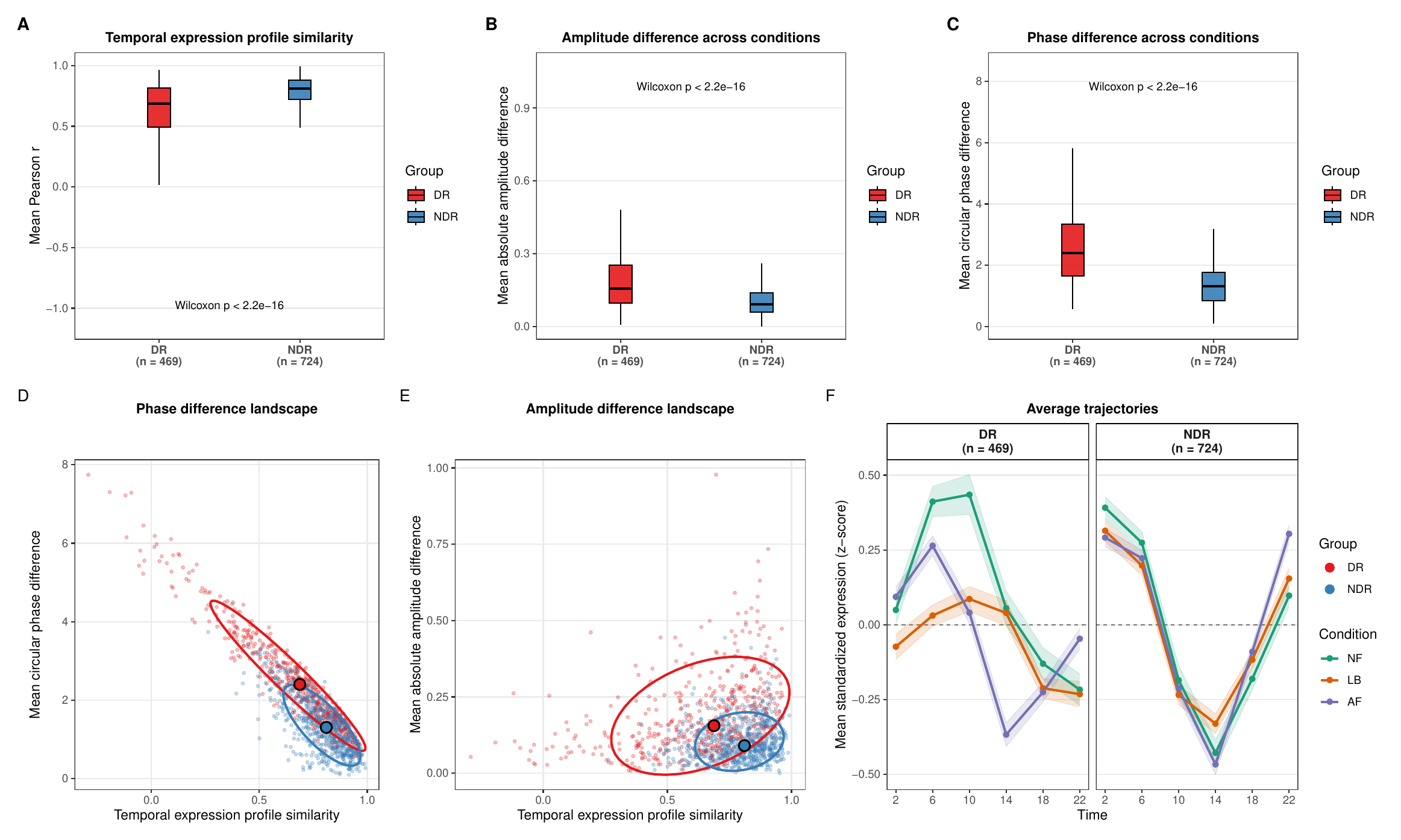}
\caption{Rhythmic expression differences of DR and NDR genes identified by CARhy-TDR}
\label{fig:DRandNDR}
\end{figure}

As shown in Figure~\ref{fig:DRandNDR}, compared to NDR genes, DR genes showed significantly lower temporal expression profile similarity across conditions (Panel A), but significantly greater amplitude  difference (Panel B) and phase difference (Panel C). In the two-dimensional landscapes, DR genes were shifted toward lower temporal expression profile similarity and higher phase difference (Panel D), as well as lower expression profile similarity and higher amplitude difference (Panel E), whereas NDR genes were more concentrated in regions of high similarity and low divergence. Consistently, average expression trajectories revealed more pronounced condition-dependent separation among DR genes, while NDR genes displayed more concordant temporal patterns across conditions (Panel F). These results indicate that DR genes exhibit greater divergence in rhythmic features across conditions, whereas NDR genes maintain relatively similar expression patterns. Additionally, of these 1,193 rhythmic genes, CARhy-TDA identified 222 genes that were differential in amplitude, and CARhy-TDP identified 506 genes that were differential in phase. Specifically, 90 genes showed differential amplitude only, 374 genes showed differential phase only, and 132 genes showed both differential amplitude and phase (Figure~\ref{fig:S5}).

\begin{figure}[!htbp]
\centering
\includegraphics[angle=90,height=0.96\textheight,keepaspectratio]{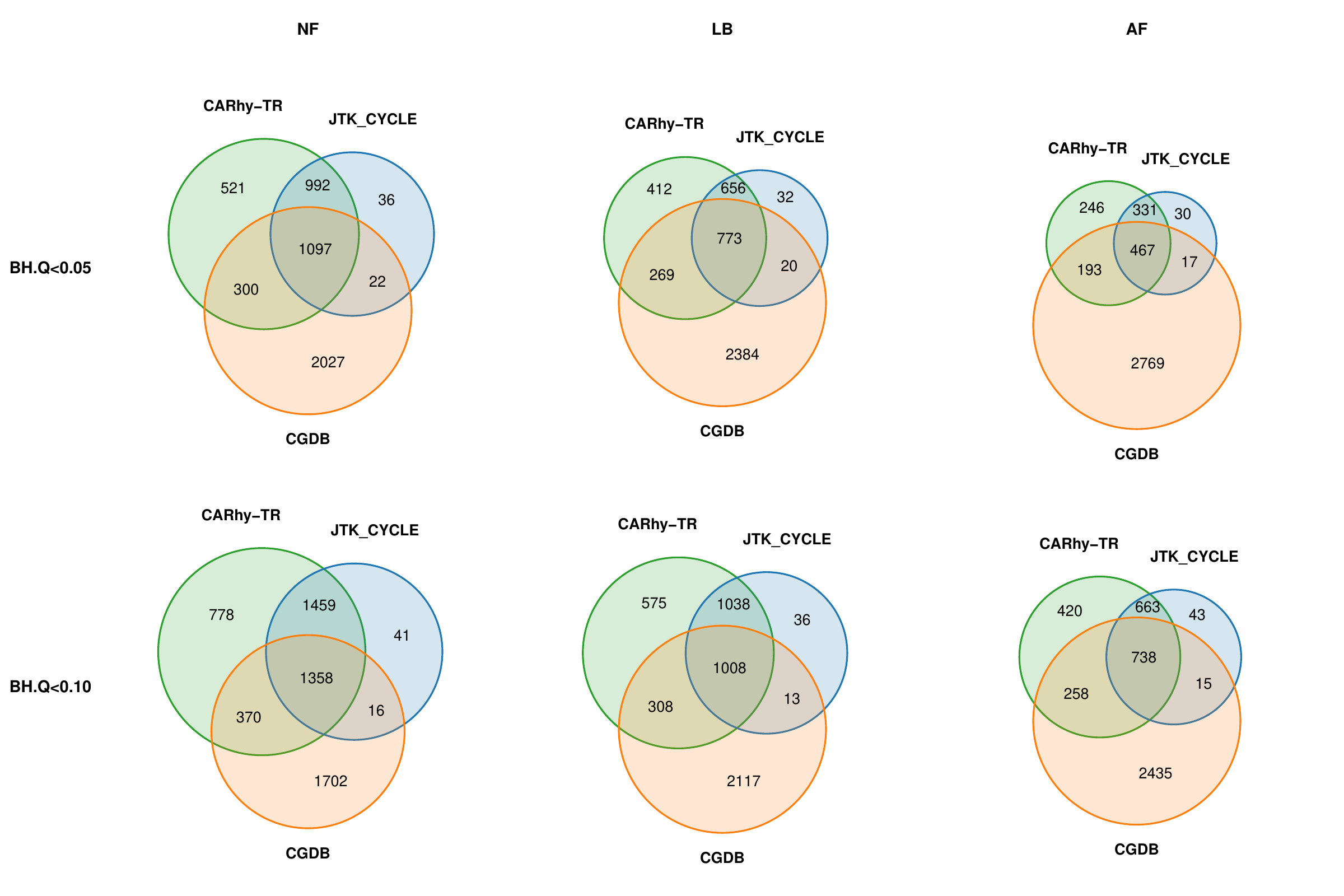}
\caption{Venn diagram showing overlaps among rhythmic genes detected by CARhy-TR and JTK\_CYCLE and the circadian gene databas (external reference set).}
\label{fig:venn}
\end{figure}

To validate our statistical framework, we compared CARhy's performance with some existing methods.  We retrieved experimentally identified rhythmic genes in mouse liver from the CGDB database (February 2026; http://cgdb.biocuckoo.org) under a light–dark design and found that 3,446 of 12,365 genes were validated as circadian genes (Table S5). We then assessed rhythmicity of the 12,365 genes for each feeding paradigm using CARhy-TR (Section \ref{sec:testrhythmicity}) and JTK\_CYCLE, respectively. The numbers of rhythmic genes detected by each method, as well as their overlap with the 3,446 rhythmic genes reported in the CGDB database are shown in the Venn diagram (Figure~\ref{fig:venn}). CARhy-TR detected rhythmicity in more of these 3,446 genes than JTK\_CYCLE did. Figure~\ref{fig:S6} shows the density of p-values from CARhy-TR and JTK\_CYCLE. CARhy-TR exhibits more enrichment of small p-values, suggesting higher detection power, whereas JTK\_CYCLE is more conservative, with p-values shifted toward larger values. This is consistent with our simulation results.

We used JTK\_CYCLE to assess rhythmicity of the 12,365 genes in NF and AF separately. For each condition, p-values across the 12,365 genes were adjusted using the Benjamini–Hochberg (BH) method at an FDR threshold of 0.05. Genes were defined as differentially rhythmic if they were rhythmic in one condition (BH.Q $<$ 0.05) but not rhythmic in the other (BH.Q $\geq$ 0.05). Using this criterion, 2,074 genes were classified as differentially rhythmic and were used as the reference set for subsequent analyses. Figure~\ref{fig:S7} illustrates the standardized temporal expression profiles of these differential rhythmic genes.

\begin{figure}[!htbp]
\centering
\includegraphics[width=1.0\textwidth]{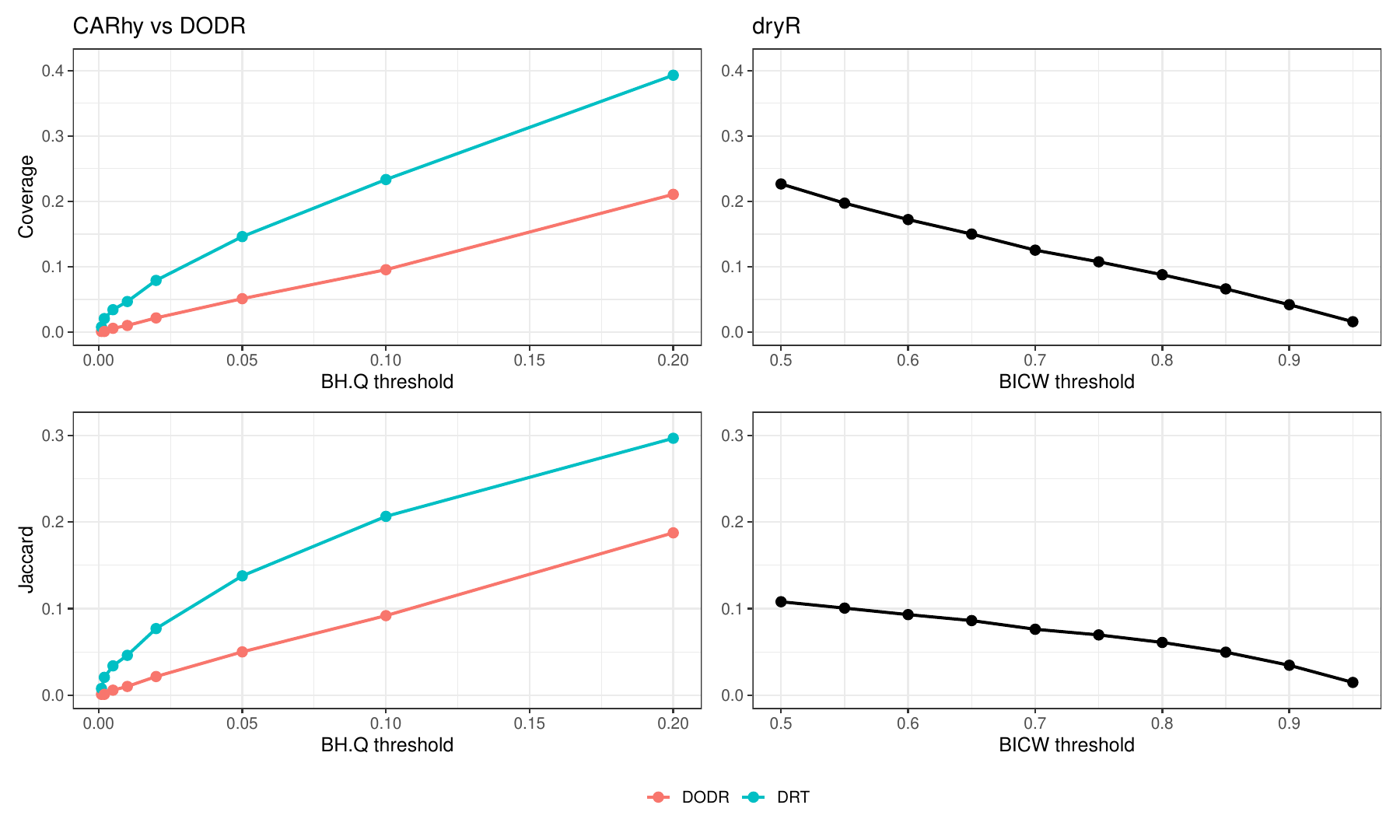}
\caption{Coverage and Jaccard index of CARhy, DODR and dryR across thresholds.} 
\label{fig:coverage_Jaccard}
\end{figure}

Using these 2,074 differential rhythmic genes as the reference set, we compared CARhy-TDR (Section \ref{AprxF}) with DODR and dryR for detecting differential rhythmicity. We compared their coverage rates and Jaccard indices across a range of P-value thresholds (and BICW thresholds for dryR). Overall, compared with DODR and dryR, CARhy-TDR achieved higher coverage rates, indicating improved recall of differential rhythmic genes, and higher Jaccard indices, indicating greater overlap with the reference set (Figure~\ref{fig:coverage_Jaccard}). Moreover, CARhy-TDR runs substantially faster than dryR (Table~\ref{tab:metrics}), and it provides raw p-values, enabling flexible post hoc selection of significance thresholds, whereas dryR requires an additional step to choose the BICW cutoff, which can be challenging to determine in practice. Compared with DODR, CARhy is not only faster but also supports experimental designs with more than two conditions and outputs rhythmic components for interpretation.

\begin{table}[!htbp]
\centering
\caption{Performance comparison among CARhy, DODR and dryR. }
\renewcommand{\arraystretch}{1.2}

\begin{tabular}{lccccccccc}
\hline
\multirow{2}{*}{metrics} & condition & discoveries & amplitude & phase & mesor & p-value & \multirow{2}{*}{runtime*} \\
& capacity & (\# genes$^{\dagger}$) & output & output & output & output  & \\
\hline
CARhy  & $\geq$ 2 & 303 & Yes & Yes & Yes & Yes & 3.7s (0.1s) \\
DODR   & 2 & 106 & No & No & No & Yes & 19.8s (1.3s) \\
dryR   & $\geq$ 2 & 33 & Yes & Yes & Yes & No & 32.3s (0.1s) \\
\hline
\end{tabular}
\vspace{1mm}

\footnotesize{ 
\begin{flushleft}
*Runtime is reported as median (IQR) elapsed seconds over 5 runs when comparing two conditions on a 4-vCPU Intel Xeon (2.20 GHz) Linux environment with 31 GiB RAM. \\ $^{\dagger}$CARhy/DODR at BH.Q=0.05; dryR at BICW=0.95.
\end{flushleft}}
\label{tab:metrics}
\end{table}

\section{Discussion}

We propose an inferential framework based on a first-harmonic Fourier regression, fitting each condition separately and testing differential rhythmicity via contrast matrices. This framework allows the error variances to differ across conditions, avoiding the homoscedasticity assumption that is often violated in practice. 
Moreover, extensive simulation studies demonstrate
excellent performance 
of the proposed methods   under unbalanced designs.

Given the limited sample sizes in typical in vivo biological experiments, we derive finite-sample approximate reference distributions for our quadratic-form statistics via a Satterthwaite type approximation, which makes use of the moment matching technique. Although approximate, the proposed tests for checking differential effects work beautifully in all different simulation scenarios. The proposed rhythmicity test uses a novel numerical technique to approximate the p-value. Our framework accommodates a general linear model with multidimensional contrasts, enabling differential rhythmicity analyses in transcriptomic studies involving multiple experimental conditions.

Most existing parametric approaches for detecting differential rhythmic genes are designed for two-condition comparisons, which limit their use in studies with more complex experimental designs. Additionally, these tools, such as DODR \citep{PMID27207944} and compareRhythms \citep{PMID34189845}, provide an overall differential rhythmicity test but do not cleanly attribute the rhythmicity change to specific rhythm components (amplitude or phase). Some tools, such as CircaCompare \citep{PMID31588519}, diffCircadian \citep{PMID34117739} and DiffCircaPipeline \citep{PMID36655766}, can test specific differences in rhythmic components, but they rely on strong noise assumptions (e.g., homogeneous error variance), which is difficult to satisfy in real omics data. Our comprehensive approach provides separate tests for overall differential rhythmicity as well as for differential amplitude and phase, and it accommodates more than two conditions without requiring homoskedasticity. Moreover, we also provide the test of rhythmicity under any experimental condition as well as the differential mesor test. All these are new in the literature.

We applied the proposed inferential framework to transcriptomic data from mouse liver. We compared CARhy's performance of detecting rhythmicity with JTK\_CYCLE, the most widely used nonparametric approach of detecting rhythmicity in transcriptomic data. CARhy achieved higher detection power, whereas JTK\_CYCLE was more conservative, consistent with our simulation results. We then evaluated CARhy for differential rhythmicity detection against two commonly used parametric methods, DODR and dryR, and found that CARhy was more powerful and identified more differentially rhythmic genes than either method, also consistent with our simulation results.

DODR's HANOVA model fits gene expression across two conditions with a single shared mesor (baseline). If the two groups have big difference in mesor, forcing them to share one baseline will cause larger residuals, which reduces the signal-to-noise ratio and thereby lowers the detection power. Weger et al. developed dryR for differential rhythmicity analysis across multiple conditions. It assigned genes to the model with maximum BICW and applied additional filters to label uncertain cases as ambiguous (e.g., “ambiguous” if the preferred model does not exceed a cutoff value (minimum BICW), with different BICW cutoffs used depending on the number of conditions, and using Cook’s distance to flag influential outliers). When comparing NF and AF, we used the recommended BICW threshold of 0.95, yet detected very few differentially rhythmic genes, indicating limited power. Simulation studies demonstrated that its statistical power of testing differential rhythmicity is highly sensitive to the chosen threshold.  Such a cutoff is subjective, does not directly correspond to a controlled error rate (e.g., FDR), and therefore is challenging for biologists to implement in practice.

The existing literature along with the  proposed method, appears to   focus on the marginal expression of the transcriptomics. 
The next phase of the research is to incorporate meaningful correlations among genes (perhaps those within the same biological pathways) when discovering biologically important markers. Extending the methodology to single-cell data, where zero counts are prevalent, also represents an important direction for future research. Additionally, a promising extension is to develop a statistical framework for the joint analysis of transcriptomic and proteomic data, which would help link gene expression changes to protein-level regulation and thereby better connect molecular alterations with functional phenotypes and disease mechanisms.

\section*{Data and Code Availability}
The experimental data are available in the Gene Expression Omnibus (GEO) under accession GSE118967 \cite{PMID30995463} and the R package for the  computation CARhy is publicly available at:\\
\url{https://github.com/DrHuang123/Comprehensive-Analyses-of-Circadian-Rhythms-CARhy}.

\section*{Supplementary information} Proofs of all theorems, as well as all tables and figures labeled with ``S'', are provided in the supplementary material. The real data analysis results referenced in paragraph 2 of Section \ref{section:4} are available in the ancillary Excel file (Table S4). The rhythmic genes in the external reference set are provided in the ancillary CSV file (Table S5).

\section*{Funding}
This research was supported by a grant from the National Heart, Lung, and Blood Institute (NHLBI) at the National Institutes of Health (NIH) (R21HL181700; to SS and JSM). Work in the lab of JSM was  also supported by grants from the National Institute of Diabetes and Digestive and Kidney Diseases (NIDDK) at NIH (R01DK128133) and the National Institute of General Medical Sciences (NIGMS) at NIH (R01GM145737).

\section*{Conflict of Interest}
The authors declare no conflict of interest.

\bibliography{references.bib}

\clearpage

\setcounter{section}{0}
\setcounter{figure}{0}
\setcounter{table}{0}
\setcounter{equation}{0}

\renewcommand{\thesection}{S\arabic{section}}
\renewcommand{\thefigure}{S\arabic{figure}}
\renewcommand{\thetable}{S\arabic{table}}
\renewcommand{\theequation}{S\arabic{equation}}

{\large\textbf{Supplementary material for ``CARhy: Comprehensive Analyses of Circadian Rhythms in Transcriptomic Experiments with Multiple Conditions"}}

\section*{Appendix}

\setcounter{equation}{0} 
\renewcommand{\theequation}{A.\arabic{equation}}
\renewcommand{\theHequation}{appendix.A.\arabic{equation}}

\setcounter{section}{0}
\renewcommand{\thesection}{A.\arabic{section}}
\renewcommand{\theHsection}{appendix.A.\arabic{section}}

\renewcommand{\thetable}{S\arabic{table}}
\renewcommand{\theHtable}{supp.table.\arabic{table}}

\renewcommand{\thefigure}{S\arabic{figure}}
\renewcommand{\theHfigure}{supp.figure.\arabic{figure}}

\section{Proof of Theorem 1}
\label{app:proof-thm1}

Under the stated assumptions, $L\wh\Gamma$
follows multivariate-Normal$(L\Gamma, \Omega)$, where 
$\Omega=L\Sigma L^\top$, and $\cov(\wh\Gamma)=\Sigma$. 
Define $A=(L\wh\Sigma L^\top)^{-1}$.
Let us calculate moments of $\rho T=(L\wh\Gamma)^\top 
AL\wh\Gamma$.  
Under $H_0$, 
$L\wh\Gamma\sim {\rm Normal}(0, \Omega)$, so, 
\begin{eqnarray*}
    E\{(L\wh\Gamma)^\top A (L\wh\Gamma)\}
=  E[\tr\{(L\wh\Gamma)^\top A (L\wh\Gamma)\}]
&\stackrel{(a)}{=}& E[\tr\{A L\wh\Gamma (L\wh\Gamma)^\top\}]\\
&=& \tr[
E\{AL\wh\Gamma (L\wh\Gamma)^\top\}]\\
&=& \tr[
E\{A E\{ L\wh\Gamma (L\wh\Gamma)^\top|A\}]\\
&\stackrel{H_0}{=}& \tr\{E(A \Omega)\}\\
&=& \tr\{E(A)\Omega)\}\\
&=& E \{\tr( \Omega A)\},
\end{eqnarray*}
where $(a)$ holds due to the cyclic property of trace.   

Recall that the random matrix 
$A=(L\wh\Sigma L^\top)^{-1}$ is a function of $\wh\sigma^2_{1}, \dots, 
\wh\sigma^2_{\mK}$ and $\Omega^{-1}=(L\Sigma L^\top)^{-1}$. 
In the following calculation, we will 
apply the second order Taylor series expansion to $A$. 
So, we shall apply the following formula for the derivative a matrix of inverse, 
$dA= d (L\wh \Sigma L^\top)^{-1}= -(L\wh \Sigma L^\top)^{-1}  
(L d\wh\Sigma L^\top) (L\wh \Sigma L^\top)^{-1}
=-A  (L d\wh\Sigma L^\top)A$ and 
\begin{eqnarray*}
   d^2A&=& d^2 (L\wh \Sigma L^\top)^{-1}\\
   &=& -d\{A  (L d\wh\Sigma L^\top)A\}\\
   &=&-dA  (L d\wh\Sigma L^\top)A
  -A  (L d^2\wh\Sigma L^\top)A
-A  (L d\wh\Sigma L^\top)dA\\
&=&A  (L d\wh\Sigma L^\top)A  (L d\wh\Sigma L^\top)A
  -A  (L d^2\wh\Sigma L^\top)A
  + A  (L d\wh\Sigma L^\top)A  (L d\wh\Sigma L^\top)A\\
  &=& 2A  (L d\wh\Sigma L^\top)A  (L d\wh\Sigma L^\top)A
  -A  (L d^2\wh\Sigma L^\top)A.
\end{eqnarray*}
Define $\sigma^2=(\sigma^2_1, \cdots, \sigma^2_\mK)$, 
$\wh\sigma^2=(\wh\sigma^2_1, \cdots, \wh\sigma^2_\mK)$.
With $\wh\Sigma_{k}=\wh\sigma^2_{k}(X^\top_k X_k)^{-1}$,
$\wh\Sigma= 
{\rm Diag}(\wh\Sigma_{1}, \dots, 
\wh\Sigma_{\mK})$, 
$
B_{1}=  L{\rm Diag}((X^\top_k X_k)^{-1}, \cdots, 0) L^\top  \Omega^{-1}, \cdots, 
B_{\mK}=  L{\rm Diag}(0, \cdots,  (X^\top_k X_k)^{-1}) L^\top  \Omega^{-1}$. Then 
\begin{eqnarray*}
   \frac{d}{d\wh\sigma^2_k}A\mid_{\wh\sigma^2=\sigma^2}&=&-\Omega^{-1} L {\rm Diag}(0, \cdots, 0, (X^\top X)^{-1}, 0, \cdots, 0) L^\top \Omega^{-1} \\
&=& -\Omega^{-1}B_k,\\
 \frac{d^2}{d(\wh\sigma^2_k)^2}A\mid_{\wh\sigma^2=\sigma^2}
 &=&2\Omega^{-1}
 L {\rm Diag}(0, \cdots, 0, (X^\top_k X_k)^{-1}, 0, \cdots, 0) L^\top \Omega^{-1}\\
 &&\times  L {\rm Diag}(0, \cdots, 0, (X^\top_k X_k)^{-1}, 0, \cdots, 0) L^\top \Omega^{-1}\\
 &&- 
\Omega^{-1} L {\rm Diag}(0, \cdots, 0, 0, 0, \cdots, 0) L^\top
 \Omega^{-1}\\
 &=& 2\Omega^{-1}B_kB_k,\\
 \frac{d^2}{d(\wh\sigma^2_k)
 d(\wh\sigma^2_{s})
}A\mid_{\wh\sigma^2=\sigma^2}
 &=&\Omega^{-1}
 L {\rm Diag}(0, \cdots, 0, (X^\top_k X_k)^{-1}, 0, \cdots, 0) L^\top \Omega^{-1}\\
 &&\times  L {\rm Diag}(0, \cdots, 0, 0, \cdots, 0, (X^\top_s X_s)^{-1}, 0, \cdots, 0) L^\top \Omega^{-1}\\
&&+\Omega^{-1}
L {\rm Diag}(0, \cdots, 0, 0, \cdots, 0, (X^\top_s X_s)^{-1}, 0, \cdots, 0) L^\top \Omega^{-1}\\
&&\times 
L {\rm Diag}(0, \cdots, 0, (X^\top_k X_k)^{-1}, 0, \cdots, 0) L^\top \Omega^{-1}\\
 &=& \Omega^{-1}B_kB_s+ \Omega^{-1}B_sB_k.
\end{eqnarray*}
Now, 
$E\left\{\tr(\Omega A)\right\}
=\mu_1+o(n^{-1})$,
where 
\begin{eqnarray*}
\mu_1&=&E\biggl(\tr\biggl[\Omega \biggl\{\Omega^{-1}
-\sum^\mK_{k=1}(\wh\sigma^2_{k}
-\sigma^2_{k})
\Omega^{-1} B_k+\sum^\mK_{k=1}\frac{(\wh\sigma^2_{k}
-\sigma^2_{k})^2}{2}2\Omega^{-1}B_kB_k\\
&& +  \sum_{s>k}(\wh\sigma^2_{k}
-\sigma^2_{k}) (\wh\sigma^2_{s}
-\sigma^2_{s}) \Omega^{-1}(B_kB_s+B_sB_k)\biggl\}\biggl]\biggl)
\\
&\stackrel{(c)}
{=}& 
\rho+  E\biggl[\tr
\biggl\{ 
\sum^\mK_{k=1}(\wh\sigma^2_{k}-\sigma^2_{k})^2
B_kB_k\biggl\}\biggl]\\
&=&\rho+ \sum^\mK_{k=1} E(\wh\sigma^2_{k}-\sigma^2_{k})^2
\tr(B_k B_k)\\
&\stackrel{(d)}{=}& \rho+ \sum^\mK_{k=1}\frac{2\sigma^4_{k}}{(n_k-3)}
\tr(B_k B_k).
\end{eqnarray*}
Equality $(c)$ follows  $\rho=\tr(\Omega\Omega^{-1})={\rm rank}(\Omega)$,  
 and 
the first order terms disappear as 
$E(\wh\sigma^2_{k})=\sigma^2_{k}$. Equality (d)
follows as $\wh\sigma^2_{k}\sim \sigma^2_{k}(n_k-3)^{-1}\chi^2_{n_k-3}$.
Next, consider 
\begin{eqnarray}
\mu_2
&=& \var\left\{ (L\wh\Gamma)^\top  A (L\wh\Gamma) \right\}\nonumber\\
&=& \var\left[ E\{(L\wh\Gamma)^\top  A (L\wh\Gamma)\mid A\} \right]+ E\left[ \var\{(L\wh\Gamma)^\top  A (L\wh\Gamma)\mid A\} \right]\nonumber\\
&=& \var\{\tr(\Omega A)\}+ 2 E\{\tr(A\Omega A\Omega)\}, \label{eq:4mu2}
\end{eqnarray}
where the last equality holds due to 
some steps considered earlier and under $H_0$, 
$L\wh \Gamma$ follows the mean zero multivariate normal distribution with covariance $\Omega$, so $\var\{ (L\wh \Gamma)^\top A
(L\wh\Gamma)\mid A\}=2\tr(A\Omega A\Omega)$. 
Now we will calculate each of the terms separately. 
First, using the second order Taylor series expansion to $A$, 
\begin{eqnarray*}
&&\var\{\tr(\Omega A)\}\\
&\approx& \var\biggl(\tr\biggl[\Omega \biggl\{\Omega^{-1}
-\sum^\mK_{k=1}(\wh\sigma^2_{k}
-\sigma^2_{k})
\Omega^{-1} B_k+\sum^\mK_{k=1}
(\wh\sigma^2_{k}
-\sigma^2_{k})^2\Omega^{-1}B_kB_k\\
&& +  \sum_{s>k}(\wh\sigma^2_{k}
-\sigma^2_{k}) (\wh\sigma^2_{s}
-\sigma^2_{s}) \Omega^{-1}(B_kB_s+B_sB_k)\biggl\}
\biggl]\biggl)
\\
&=&
\var\biggl(\tr\biggl[ \biggl\{I
-\sum^\mK_{k=1}(\wh\sigma^2_{k}
-\sigma^2_{k})
 B_k+\sum^\mK_{k=1}
(\wh\sigma^2_{k}
-\sigma^2_{k})^2B_kB_k\\
&& +  \sum_{s>k}(\wh\sigma^2_{k}
-\sigma^2_{k}) (\wh\sigma^2_{s}
-\sigma^2_{s}) (B_kB_s+B_sB_k)\biggl\}
\biggl]\biggl)
\\
&=& \sum^\mK_{k=1}\biggl(\var(\wh\sigma^2_{k})\{\trace(B_k)\}^2
+
\{\trace(B_kB_k)\}^2\biggl[
E\{(\wh\sigma^2_{k}-\sigma^2_{k})^4\}- \var^2(\wh\sigma^2_{k})
\biggl]\\
&& -2\trace(B_k)\trace(B_kB_k)
E\{(\wh\sigma^2_{k}-\sigma^2_{k})^3\}
\biggl)\\
&&+ \sum_{s>k}\var(\wh\sigma^2_k)\var(\wh\sigma^2_s) \{\tr(B_kB_s+B_sB_k)\}^2.  
\end{eqnarray*}
Since $\wh\sigma^2_{k}\sim \sigma^2_{k}(n_k-3)^{-1}\chi^2_{n_k-3}$, 
$\var(\wh\sigma^2_{k})=\sigma^4_{k}(n_k-3)^{-2}\times 2 (n_k-3)=2\sigma^4_{k}(n_k-3)^{-1}$, 
$E(\wh\sigma^2_{k}-\sigma^2_{k})^3=\sigma^6_{k}(n_k-3)^{-3}\times E
\{\chi^2_{n_k-3}- E(\chi^2_{n_k-3})\}^3$ and 
$E(\wh\sigma^2_{k}-\sigma^2_{k})^4=\sigma^8_{k}(n_k-3)^{-4}\times E
\{\chi^2_{n_k-3}- E(\chi^2_{n_k-3})\}^4$.
For a $\chi^2$ random variable with degrees of freedom  $\scripty{r}$, 
the $s$th order raw moment is 
$\mu^{'}_s= 2^s\Gamma{(\scripty{r}/2+s)}/\Gamma{(\scripty{r}/2)}$. 
So,
$\mu^{'}_1=\scripty{r}$, $\mu^{'}_2= 4(\scripty{r}/2+1)\scripty{r}/2$,
$\mu^{'}_3= 8 (\scripty{r}/2+2)(\scripty{r}/2+1)\scripty{r}/2$, 
$\mu^{'}_4= 16 (\scripty{r}/2+3)(\scripty{r}/2+2)(\scripty{r}/2+1)\scripty{r}/2$, and these can be used to compute 
$E\{\chi^2- E(\chi^2)\}^3= \mu^{'}_3-3\mu^{'}_2\mu^{'}_1+2
(\mu^{'}_1)^3$ and 
$E\{\chi^2- E(\chi^2)\}^4= \mu^{'}_4-4\mu^{'}_3\mu^{'}_1
+6\mu^{'}_2(\mu^{'}_1)^2
-4\mu^{'}_1(\mu^{'}_1)^3
+ (\mu^{'}_1)^4
= \mu^{'}_4-4\mu^{'}_3\mu^{'}_1
+6\mu^{'}_2(\mu^{'}_1)^2
-3(\mu^{'}_1)^4
$.

Next, again using the Taylor's expansion to $A$, 
\begin{eqnarray*}
&& E\{\tr(A\Omega A\Omega)\}\\
&\approx& 
 E 
\biggl(\tr\biggl[
\biggl\{\Omega^{-1}
-\sum^\mK_{k=1}(\wh\sigma^2_{k}
-\sigma^2_{k})
\Omega^{-1} B_k+\sum^\mK_{k=1}
(\wh\sigma^2_{k}
-\sigma^2_{k})^2\Omega^{-1}B_kB_k\\
&& +  \sum_{s>k}(\wh\sigma^2_{k}
-\sigma^2_{k}) (\wh\sigma^2_{s}
-\sigma^2_{s}) \Omega^{-1}(B_kB_s+B_sB_k)\biggl\}\Omega \\
&& 
\times \biggl\{\Omega^{-1}
-\sum^\mK_{k=1}(\wh\sigma^2_{k}
-\sigma^2_{k})
\Omega^{-1} B_k+\sum^\mK_{k=1}
(\wh\sigma^2_{k}
-\sigma^2_{k})^2\Omega^{-1}B_kB_k\\
&& +  \sum_{s>k}(\wh\sigma^2_{k}
-\sigma^2_{k}) (\wh\sigma^2_{s}
-\sigma^2_{s}) \Omega^{-1}(B_kB_s+B_sB_k)\biggl\}\Omega\biggl]\biggl) \\
&=& E 
\biggl(\tr\biggl[
\biggl\{I
-\sum^\mK_{k=1}(\wh\sigma^2_{k}
-\sigma^2_{k})
\Omega^{-1} B_k\Omega+\sum^\mK_{k=1}
(\wh\sigma^2_{k}
-\sigma^2_{k})^2\Omega^{-1}B_kB_k\Omega\\
&& +  \sum_{s>k}(\wh\sigma^2_{k}
-\sigma^2_{k}) (\wh\sigma^2_{s}
-\sigma^2_{s}) \Omega^{-1}(B_kB_s+B_sB_k)\Omega\biggl\} \\
&& 
\times \biggl\{I
-\sum^\mK_{k=1}(\wh\sigma^2_{k}
-\sigma^2_{k})
\Omega^{-1} B_k\Omega+\sum^\mK_{k=1}
(\wh\sigma^2_{k}
-\sigma^2_{k})^2\Omega^{-1}B_kB_k\Omega\\
&& +  \sum_{s>k}(\wh\sigma^2_{k}
-\sigma^2_{k}) (\wh\sigma^2_{s}
-\sigma^2_{s}) \Omega^{-1}(B_kB_s+B_sB_k)\Omega\biggl\}\biggl]\biggl) \\
&=& E 
\biggl(\tr\biggl[
I
-2\sum^\mK_{k=1}(\wh\sigma^2_{k}
-\sigma^2_{k})
\Omega^{-1} B_k\Omega+2\sum^\mK_{k=1}
(\wh\sigma^2_{k}
-\sigma^2_{k})^2\Omega^{-1}B_kB_k\Omega\\
&& +  2\sum_{s>k}(\wh\sigma^2_{k}
-\sigma^2_{k}) (\wh\sigma^2_{s}
-\sigma^2_{s}) \Omega^{-1}(B_kB_s+B_sB_k)\Omega \\
&&+ \{\sum^\mK_{k=1}(\wh\sigma^2_{k}
-\sigma^2_{k})
\Omega^{-1} B_k\Omega\}
\{\sum^\mK_{k=1}(\wh\sigma^2_{k}
-\sigma^2_{k})
\Omega^{-1} B_k\Omega\}\\
&& + \{\sum^\mK_{k=1}
(\wh\sigma^2_{k}
-\sigma^2_{k})^2\Omega^{-1}B_kB_k\Omega 
\} \{\sum^\mK_{k=1}
(\wh\sigma^2_{k}
-\sigma^2_{k})^2\Omega^{-1}B_kB_k\Omega 
\} \\
&&+\{\sum_{s>k}(\wh\sigma^2_{k}
-\sigma^2_{k}) (\wh\sigma^2_{s}
-\sigma^2_{s}) \Omega^{-1}(B_kB_s+B_sB_k)\Omega\} \\
&&\hskip 5mm \times 
\{
\sum_{s>k}(\wh\sigma^2_{k}
-\sigma^2_{k}) (\wh\sigma^2_{s}
-\sigma^2_{s}) \Omega^{-1}(B_kB_s+B_sB_k)\Omega
\}
\\
&& - \{\sum^\mK_{k=1}(\wh\sigma^2_{k}
-\sigma^2_{k})
\Omega^{-1} B_k\Omega\}\{\sum^\mK_{k=1}
(\wh\sigma^2_{k}
-\sigma^2_{k})^2\Omega^{-1}B_kB_k\Omega 
\}\\
&& - \{\sum^\mK_{k=1}(\wh\sigma^2_{k}
-\sigma^2_{k})
\Omega^{-1} B_k\Omega\}\{
\sum_{s>k}(\wh\sigma^2_{k}
-\sigma^2_{k}) (\wh\sigma^2_{s}
-\sigma^2_{s}) \Omega^{-1}(B_kB_s+B_sB_k)\Omega
\}\\
&&  - \{\sum^\mK_{k=1}
(\wh\sigma^2_{k}
-\sigma^2_{k})^2\Omega^{-1}B_kB_k\Omega 
\}\{\sum^\mK_{k=1}(\wh\sigma^2_{k}
-\sigma^2_{k})
\Omega^{-1} B_k\Omega\}\\
&&- \{
\sum_{s>k}(\wh\sigma^2_{k}
-\sigma^2_{k}) (\wh\sigma^2_{s}
-\sigma^2_{s}) \Omega^{-1}(B_kB_s+B_sB_k)\Omega
\}\{\sum^\mK_{k=1}(\wh\sigma^2_{k}
-\sigma^2_{k})
\Omega^{-1} B_k\Omega\}\\
&& +\{\sum^\mK_{k=1}
(\wh\sigma^2_{k}
-\sigma^2_{k})^2\Omega^{-1}B_kB_k\Omega 
\} \{
\sum_{s>k}(\wh\sigma^2_{k}
-\sigma^2_{k}) (\wh\sigma^2_{s}
-\sigma^2_{s}) \Omega^{-1}(B_kB_s+B_sB_k)\Omega
\}\\
&& +  \{
\sum_{s>k}(\wh\sigma^2_{k}
-\sigma^2_{k}) (\wh\sigma^2_{s}
-\sigma^2_{s}) \Omega^{-1}(B_kB_s+B_sB_k)\Omega
\}
\{\sum^\mK_{k=1}
(\wh\sigma^2_{k}
-\sigma^2_{k})^2\Omega^{-1}B_kB_k\Omega 
\}\biggl]\biggl).
\end{eqnarray*}
 Next, we will calculate trace and expectations and we will omit the terms whose expectation is zero. So, 
 \begin{eqnarray*}
  &&E\{ \tr(A\Omega A\Omega)\}   \\
&\approx & \rho+ 2 \sum^\mK_{k=1} \var(\wh\sigma^2_k)\tr(\Omega^{-1}B_kB_k\Omega)\\
&& + \sum^\mK_{k=1} \var(\wh\sigma^2_k)\tr(\Omega^{-1}B_k \Omega  \Omega^{-1}B_k \Omega)\\
&& +\sum^\mK_{k=1} E(\wh\sigma^2_k-\sigma^2_k)^4\tr(\Omega^{-1}B_kB_k \Omega  \Omega^{-1}B_kB_k \Omega)\\
&&+ 2\sum_{s>k} \var(\wh\sigma^2_k)\var(\wh\sigma^2_s)\tr(\Omega^{-1}B_kB_k \Omega  \Omega^{-1}B_sB_s \Omega)\\
 &&+ \sum_{s>k} \var(\wh\sigma^2_k)\var(\wh\sigma^2_s)
 \tr\{\Omega^{-1}(B_kB_s+B_sB_k)\Omega \Omega^{-1}(B_kB_s+B_sB_k)\Omega\}\\
&&- \sum^\mK_{k=1} E(\wh\sigma^2_k-\sigma^2_k)^3\tr(
\Omega^{-1}B_k\Omega \Omega^{-1} B_kB_k \Omega
)\\
&& - \sum^\mK_{k=1} E(\wh\sigma^2_k-\sigma^2_k)^3 \tr(
\Omega^{-1}B_kB_k \Omega \Omega^{-1} B_k\Omega)\\
&= & \rho+ 2 \sum^\mK_{k=1} \var(\wh\sigma^2_k)\tr(B_kB_k)\\
&& + \sum^\mK_{k=1} \var(\wh\sigma^2_k)\tr(B_kB_k)\\
&& +\sum^\mK_{k=1} E(\wh\sigma^2_k-\sigma^2_k)^4\tr(B_kB_k B_kB_k )\\
&&+ 2\sum_{s>k} \var(\wh\sigma^2_k)\var(\wh\sigma^2_s)\tr(B_kB_kB_sB_s)\\
 &&+ \sum_{s>k} \var(\wh\sigma^2_k)\var(\wh\sigma^2_s)
 \tr\{(B_kB_s+B_sB_k)(B_kB_s+B_sB_k)\}\\
&&- \sum^\mK_{k=1} E(\wh\sigma^2_k-\sigma^2_k)^3\tr(
B_k B_kB_k 
)\\
&& - \sum^\mK_{k=1} E(\wh\sigma^2_k-\sigma^2_k)^3 \tr(
B_kB_k B_k)\\
&=&  \rho+ 3 \sum^\mK_{k=1} \var(\wh\sigma^2_k)\tr(B_kB_k)\\
&& +\sum^\mK_{k=1} E(\wh\sigma^2_k-\sigma^2_k)^4\tr(B_kB_k B_kB_k )\\
&&+ \sum_{s>k} \var(\wh\sigma^2_k)\var(\wh\sigma^2_s)\{2\tr(B_kB_kB_sB_s)+ 
 \tr\{(B_kB_s+B_sB_k)(B_kB_s+B_sB_k)\}\\
&&- 2\sum^\mK_{k=1} E(\wh\sigma^2_k-\sigma^2_k)^3\tr(
B_k B_kB_k 
). 
\end{eqnarray*}
Now, placing all the terms in (\ref{eq:4mu2}),  we obtain 
\begin{eqnarray*}
 \mu_2&=&  \sum^\mK_{k=1}\biggl(\var(\wh\sigma^2_{k})\{\trace(B_k)\}^2
+
\{\trace(B_kB_k)\}^2\biggl[
E\{(\wh\sigma^2_{k}-\sigma^2_{k})^4\}- \var^2(\wh\sigma^2_{k})
\biggl]\\
&& -2\trace(B_k)\trace(B_kB_k)
E\{(\wh\sigma^2_{k}-\sigma^2_{k})^3\}
\biggl)\\
&&+ \sum_{s>k}\var(\wh\sigma^2_k)\var(\wh\sigma^2_s) \{\tr(B_kB_s+B_sB_k)\}^2
\\
&&
 + 2\rho+ 6 \sum^\mK_{k=1} \var(\wh\sigma^2_k)\tr(B_kB_k)\\
&& +2\sum^\mK_{k=1} E(\wh\sigma^2_k-\sigma^2_k)^4\tr(B_kB_k B_kB_k )\\
&&+ 2\sum_{s>k} \var(\wh\sigma^2_k)\var(\wh\sigma^2_s)\{2\tr(B_kB_kB_sB_s)+ 
 \tr\{(B_kB_s+B_sB_k)(B_kB_s+B_sB_k)\}\\
&&- 4\sum^\mK_{k=1} E(\wh\sigma^2_k-\sigma^2_k)^3\tr(
B_k B_kB_k 
)\\
&=&2\rho\\
&&+\sum^\mK_{k=1} \var(\wh\sigma^2_{k})\biggl[
\{\trace(B_k)\}^2+ 6\tr(B_kB_k)\biggl]\\
 &&-\sum^\mK_{k=1}
E\{(\wh\sigma^2_{k}-\sigma^2_{k})^3\}\biggl
\{2\trace(B_k)\tr(B_kB_k)   +4 \tr(B_kB_kB_k)\biggl\}
\\
&&+  \sum^\mK_{k=1}
E\{(\wh\sigma^2_{k}-\sigma^2_{k})^4\}\biggl[
\{\trace(B_kB_k)\}^2   +2 \tr(B_kB_kB_kB_k)\biggl]\\
&&-\sum^\mK_{k=1} \var^2(\wh\sigma^2_{k})\{\tr(B_kB_k)\}^2\\
&& +\sum_{s>k} \var(\wh\sigma^2_k)\var(\wh\sigma^2_s)\biggl[
\{\tr(B_kB_s+B_sB_k)\}^2+4\tr(B_kB_kB_sB_s)\\
&&\hskip 5mm + 
 2\tr\{(B_kB_s+B_sB_k)(B_kB_s+B_sB_k)\biggl].
\end{eqnarray*}

Assuming $cT= c(L\wh\Gamma_g)^\top A (L\wh\Gamma_g)/\rho$ with 
$c$ being a constant and $\rho$ being the rank of $L$, 
follows $F_{\rho, df}$,
its theoretical first moment and variance are $df/(df-2)$ and
$2df^2(df+\rho-2)/ \rho(df-2)^2(df-4)$, respectively. 
Now, equating the moments we have two equations
\begin{eqnarray*}
\frac{df}{df-2}&=& \frac{c\mu_1}{\rho},\\
\frac{2df^2(df+\rho-2)}{\rho(df-2)^2(df-4)}&=& \frac{c^2\mu_2}{\rho^2},
\end{eqnarray*}
and solving these two we obtain the following
\begin{eqnarray*}
    df=\frac{\rho-2 +2\rho \mu_2/\mu^2_1}{\rho\mu_2/2\mu^2_1-1},
\end{eqnarray*}
and 
\begin{eqnarray*}
    c= \frac{\rho df}{\mu_1(df-2)}.
\end{eqnarray*}

\section{Proof of Theorem 2}
\label{app:proof-thm2}

Under the stated assumptions,  
$L\wh\theta$
follows multivariate-Normal$(L\theta, \Psi)$,
where the covariance $\Psi=L {\rm Diag}(D^\top_1\Sigma_1D_1, \cdots, D^\top_\mK\Sigma_\mK D_\mK) L^\top$. 
Under $H_0$ (no differential amplitude or no differential phase depending on the hypothesis), $L\theta=0$, and $L\wh\theta\sim \mbox{multivariate-Normal}(0, \Psi)$.

Define $\Lambda=\{L {\rm Diag}(D^\top_1\wh\Sigma_1D_1, \cdots, D^\top_\mK\wh\Sigma_\mK D_\mK) L^\top\} ^{-1}$.
Let us calculate the first two moments of $(L\wh\theta)^\top 
\Lambda (L\wh\theta)$, which is $\rho T$.  
With the mean-zero multivariate normal distribution for $L\wh\theta$ under $H_0$,  
\begin{eqnarray*}
    E\{(L\wh\theta)^\top \Lambda (L\wh\theta)\}= E \{\tr( \Psi \Lambda)\}. 
\end{eqnarray*}
Recall that the random matrix 
$\Lambda$ is a function of $\wh\sigma^2_{1}, \dots, 
\wh\sigma^2_{\mK}$, and we obtain 
$\Psi^{-1}$ when $\wh\sigma^2_{1}, \dots, 
\wh\sigma^2_{\mK}$ are replaced with $\sigma^2_{1}, \dots, 
\sigma^2_{\mK}$ in $\Lambda$. 
As in Theorem 1, we 
apply the second order Taylor series expansion to $\Lambda$. 
So, we shall apply the following formula for the derivative a matrix of inverse, 
\begin{eqnarray*}
  d\Lambda&=& d \{L {\rm Diag}(D^\top_1\wh\Sigma_1D_1, \cdots, D^\top_\mK\wh\Sigma_\mK D_\mK) L^\top\}^{-1}\\
  &=& -\Lambda   
\{L d{\rm Diag}(D^\top_1\wh\Sigma_1D_1, \cdots, D^\top_\mK\wh\Sigma_\mK D_\mK) L^\top\} \Lambda  
\end{eqnarray*}
 and 
\begin{eqnarray*}
   d^2\Lambda 
   &=& -d[\Lambda   
\{L d{\rm Diag}(D^\top_1\wh\Sigma_1D_1, \cdots, D^\top_\mK\wh\Sigma_\mK D_\mK) L^\top\} \Lambda ]\\
   &=&-d\Lambda   \{L d{\rm Diag}(D^\top_1\wh\Sigma_1D_1, \cdots, D^\top_\mK\wh\Sigma_\mK D_\mK) L^\top\}\Lambda
  \\
  &&-\Lambda  \{L d^2{\rm Diag}(D^\top_1\wh\Sigma_1D_1, \cdots, D^\top_\mK\wh\Sigma_\mK D_\mK) L^\top\}\Lambda\\
  && 
-\Lambda  \{L d{\rm Diag}(D^\top_1\wh\Sigma_1D_1, \cdots, D^\top_\mK\wh\Sigma_\mK D_\mK) L^\top\}d\Lambda\\
&=&\Lambda   \{L d{\rm Diag}(D^\top_1\wh\Sigma_1D_1, \cdots, D^\top_\mK\wh\Sigma_\mK D_\mK) L^\top\}\Lambda  \{L d{\rm Diag}(D^\top_1\wh\Sigma_1D_1, \cdots, D^\top_\mK\wh\Sigma_\mK D_\mK) L^\top\}\Lambda
  \\
  &&-\Lambda  \{L d^2{\rm Diag}(D^\top_1\wh\Sigma_1D_1, \cdots, D^\top_\mK\wh\Sigma_\mK D_\mK) L^\top\}\Lambda\\
  && 
+\Lambda  \{L d{\rm Diag}(D^\top_1\wh\Sigma_1D_1, \cdots, D^\top_\mK\wh\Sigma_\mK D_\mK) L^\top\}\Lambda   \{L d{\rm Diag}(D^\top_1\wh\Sigma_1D_1, \cdots, D^\top_\mK\wh\Sigma_\mK D_\mK) L^\top\}\Lambda\\
  &=& 2\Lambda  \{L d{\rm Diag}(D^\top_1\wh\Sigma_1D_1, \cdots, D^\top_\mK\wh\Sigma_\mK D_\mK) L^\top\}\Lambda   \{L d{\rm Diag}(D^\top_1\wh\Sigma_1D_1, \cdots, D^\top_\mK\wh\Sigma_\mK D_\mK) L^\top\}\Lambda\\
 && -\Lambda  \{L d^2{\rm Diag}(D^\top_1\wh\Sigma_1D_1, \cdots, D^\top_\mK\wh\Sigma_\mK D_\mK) L^\top\}\Lambda.
\end{eqnarray*}
Now, 
\begin{eqnarray*}
   \frac{d}{d\wh\sigma^2_k}\Lambda \mid_{\wh\sigma^2=\sigma^2}&=&-\Psi^{-1} L {\rm Diag}(0, \cdots, 0, D^\top_k(X^\top_k X_k)^{-1}D_k, 0, \cdots, 0) L^\top \Psi^{-1} \\
&=& -\Psi^{-1}G_k,\\
 \frac{d^2}{d(\wh\sigma^2_k)^2}\Lambda \mid_{\wh\sigma^2=\sigma^2}
 &=&2\Psi^{-1}
 L {\rm Diag}(0, \cdots, 0, D^\top_k(X^\top_k X_k)^{-1}D_k, 0, \cdots, 0) L^\top \Psi^{-1}\\
 &&\times  L {\rm Diag}(0, \cdots, 0, D^\top_k(X^\top_k X_k)^{-1}D_k, 0, \cdots, 0) L^\top \Psi^{-1}\\
 &&- 
\Psi^{-1} L {\rm Diag}(0, \cdots, 0, 0, 0, \cdots, 0) L^\top
 \Psi^{-1}\\
 &=& 2\Psi^{-1}G_kG_k,\\
 \frac{d^2}{d(\wh\sigma^2_k)
 d(\wh\sigma^2_{s})
}\Lambda \mid_{\wh\sigma^2=\sigma^2}
 &=&\Psi^{-1}
 L {\rm Diag}(0, \cdots, 0, D^\top_k(X^\top_k X_k)^{-1}D_k, 0, \cdots, 0) L^\top \Psi^{-1}\\
 &&\times  L {\rm Diag}(0, \cdots, 0, 0, \cdots, 0, D^\top_s(X^\top_s X_s)^{-1}D_s, 0, \cdots, 0) L^\top \Psi^{-1}\\
&&+\Psi^{-1}
L {\rm Diag}(0, \cdots, 0, 0, \cdots, 0, D^\top_s(X^\top_s X_s)^{-1}D_s, 0, \cdots, 0) L^\top \Psi^{-1}\\
&&\times 
L {\rm Diag}(0, \cdots, 0, D^\top_k(X^\top_k X_k)^{-1}D_k, 0, \cdots, 0) L^\top \Psi^{-1}\\
 &=& \Psi^{-1}(G_kG_s+G_sG_k).
\end{eqnarray*}
Calculation of $\mu_1$ and $\mu_2$ follow the same steps as in Theorem 1.

\begin{table}[H]
\centering
\caption{Simulation scenarios for testing  differential mesor.}\label{tab:S1}
\renewcommand{\arraystretch}{1.15}
\setlength{\tabcolsep}{6pt}

\begin{tabular}{l c c c c
  >{\centering\arraybackslash}p{1.2cm}
  >{\centering\arraybackslash}p{1.2cm}
  >{\centering\arraybackslash}p{1.2cm}}
\toprule
\multirow{2}{*}{Conditions} & \multirow{2}{*}{Case}
& \multicolumn{3}{c}{Mesor} 
& \multicolumn{3}{c}{Noise (normal dist.)}\\
\cmidrule(lr){3-5}\cmidrule(lr){6-8}
& & $\delta_1$ & $\delta_2$ & $\delta_3$
  & $\sigma_1$ & $\sigma_2$ & $\sigma_3$\\
\midrule

\multirow{2}{*}{3 conditions}
& 1 & 1 & 1 & 1  & 1 & 1 & 1 \\
& 2 & 1 & 1 & 1  & 0.5 & 2 & 2 \\
& 3 & 1 & 3 & 5  & 1 & 1 & 1 \\
& 4 & 1 & 3 & 5  & 0.5 & 2 & 2 \\
\midrule

\multirow{2}{*}{2 conditions}
& 5 & 1 & 1 & -- & 1 & 1 & -- \\
& 6 & 1 & 1 & -- & 0.5 & 2 & -- \\
& 7 & 1 & 3 & -- & 1 & 1 & -- \\
& 8 & 1 & 3 & -- & 0.5 & 2 & -- \\
\bottomrule
\end{tabular}
\end{table}

\begin{table}[H]
\centering
\caption{Simulation scenarios for testing differential amplitude.}
\label{tab:S2}
\renewcommand{\arraystretch}{1.15}
\setlength{\tabcolsep}{6pt}

\begin{tabular}{ll ccc ccc
  >{\centering\arraybackslash}p{1cm}
  >{\centering\arraybackslash}p{1cm}
  >{\centering\arraybackslash}p{1cm}}
\toprule
\multirow{2}{*}{Model} & \multirow{2}{*}{Case}
& \multicolumn{3}{c}{Amplitude}
& \multicolumn{3}{c}{Phase}
& \multicolumn{3}{c}{Noise (normal dist.)} \\
\cmidrule(lr){3-5}\cmidrule(lr){6-8}\cmidrule(lr){9-11}
& & $A_1$ & $A_2$ & $A_3$
  & $\phi_1$ & $\phi_2$ & $\phi_3$
  & $\sigma_1$ & $\sigma_2$ & $\sigma_3$ \\
\midrule

\multirow{4}{*}{3-conditions}
& 1 & 4 & 4 & 4  & 5 & 5 & 5  & 1   & 1 & 1 \\
& 2 & 4 & 4 & 4  & 5 & 5 & 5  & 0.5 & 2 & 2 \\
& 3 & 2 & 2 & 4  & 5 & 5 & 5  & 1   & 1 & 1 \\
& 4 & 2 & 2 & 4  & 5 & 5 & 5  & 0.5 & 2 & 2 \\
\midrule

\multirow{4}{*}{2-conditions}
& 5 & 4 & 4 & -- & 5 & 5 & -- & 1   & 1 & -- \\
& 6 & 4 & 4 & -- & 5 & 5 & -- & 0.5 & 2 & -- \\
& 7 & 2 & 4 & -- & 5 & 5 & -- & 1   & 1 & -- \\
& 8 & 2 & 4 & -- & 5 & 5 & -- & 0.5 & 2 & -- \\
\bottomrule
\end{tabular}
\end{table}

\begin{table}[H]
\centering
\caption{Simulation scenarios for testing differential phase.}
\label{tab:S3}
\begin{tabular}{ll ccc ccc
  >{\centering\arraybackslash}p{1cm}
  >{\centering\arraybackslash}p{1cm}
  >{\centering\arraybackslash}p{1cm}}
\toprule
\multirow{2}{*}{Model} & \multirow{2}{*}{Case}
& \multicolumn{3}{c}{Amplitude}
& \multicolumn{3}{c}{Phase}
& \multicolumn{3}{c}{Noise (normal dist.)} \\
\cmidrule(lr){3-5}\cmidrule(lr){6-8}\cmidrule(lr){9-11}
& & $A_1$ & $A_2$ & $A_3$ & $\phi_1$ & $\phi_2$ & $\phi_3$ & $\sigma_1$ & $\sigma_2$ & $\sigma_3$ \\
\midrule

\multirow{4}{*}{3-conditions}
& 9 & 4 & 4 & 4 & 5 & 5 & 5 & 1   & 1 & 1 \\
& 10 & 4 & 4 & 4 & 5 & 5 & 5 & 0.5 & 2 & 2 \\
& 11 & 4 & 4 & 4 & 5 & 3 & 5 & 1   & 1 & 1 \\
& 12 & 4 & 4 & 4 & 5 & 3 & 5 & 0.5 & 2 & 2 \\
\midrule

\multirow{4}{*}{2-conditions}
& 13 & 4 & 4 & -- & 5 & 5 & -- & 1   & 1 & -- \\
& 14 & 4 & 4 & -- & 5 & 5 & -- & 0.5 & 2 & -- \\
& 15 & 4 & 4 & -- & 5 & 3 & -- & 1   & 1 & -- \\
& 16 & 4 & 4 & -- & 5 & 3 & -- & 0.5 & 2 & -- \\
\bottomrule
\end{tabular}
\end{table}

\medskip

Table S4: Summary of CARhy statistical results for genes in mouse live.\\
This table is provided as a separate Excel file and contains two sheets: \textit{CARhy\_results}, which reports the p values from all statistical tests implemented in CARhy for each gene, together with the estimated mesor, amplitude, and phase under each condition; and \textit{Column\_descriptions}, which provides explanations for all columns included in the Results sheet.

\medskip

Table S5: Experimentally identified rhythmic genes in mouse live from CGDB.\\
This table is provided as a separate CSV file and contains the gene symbols of the 3,446 genes validated as circadian genes in CGDB database, based on evidence from RT-PCR, Northern blot, in situ hybridization, and microarray or RNA-sequencing studies.

\newpage

\begin{figure}[H]
\centering
\includegraphics[width=1.0\textwidth]{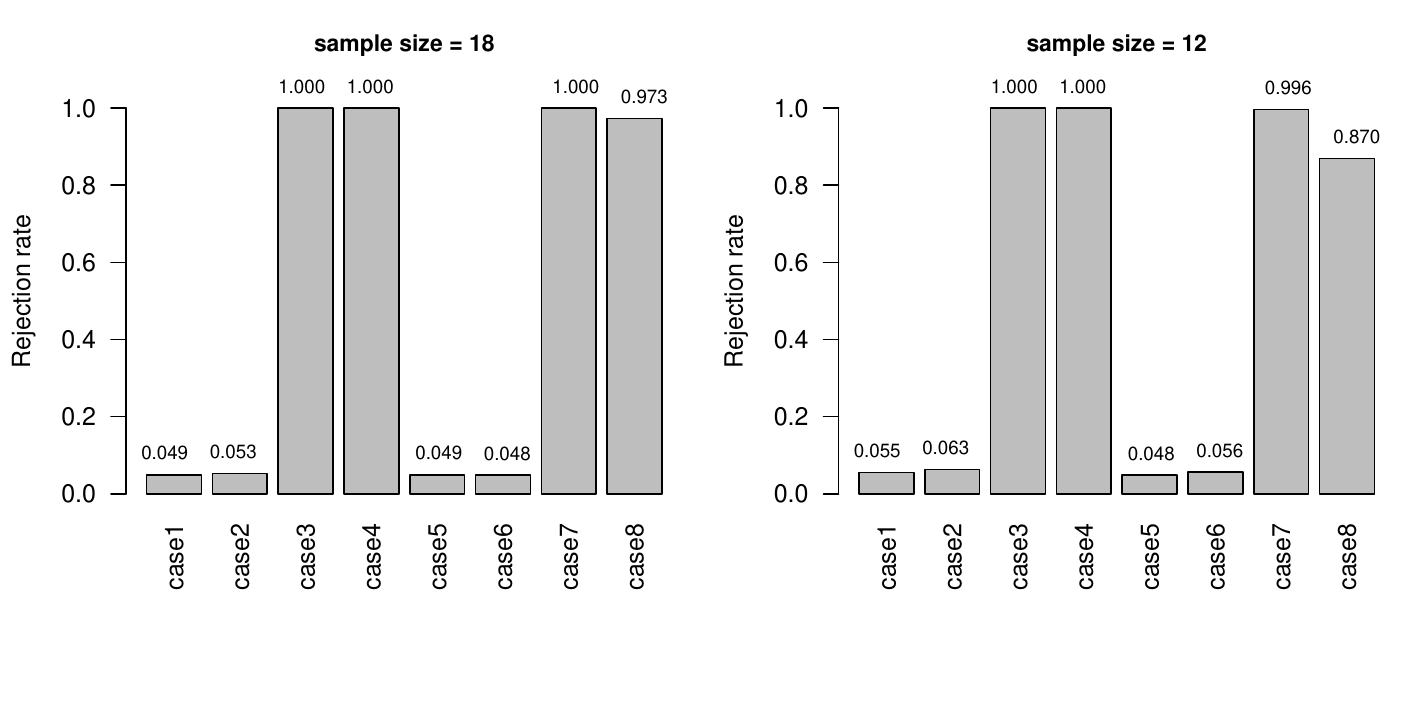}
\caption{Type-I error rates and power for testing differential mesor (Table \ref{tab:S1}).}
\label{fig:S2}
\end{figure}

\begin{figure}[H]
\centering
\includegraphics[width=1.0\textwidth]{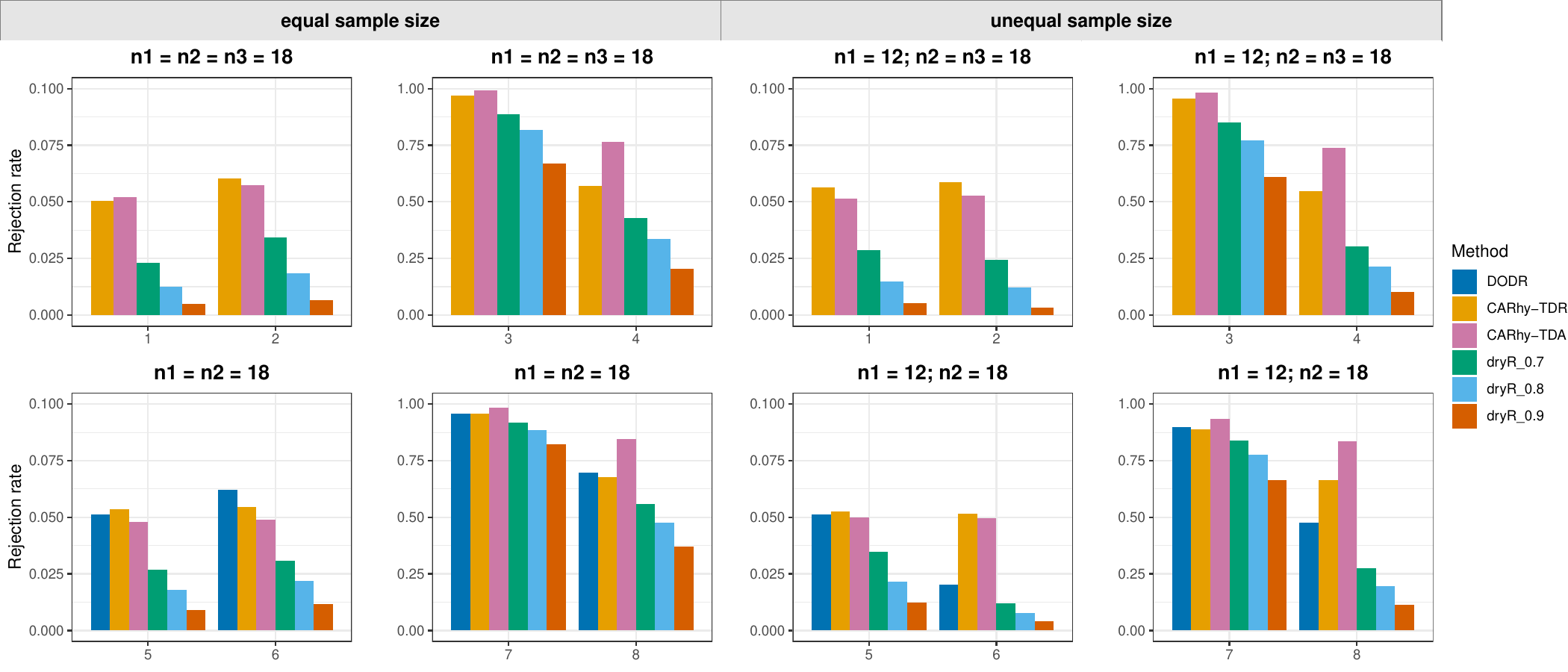}
\caption{Type-I error rate and power for testing differential amplitude (Table \ref{tab:S2}).}
\label{fig:S3}
\end{figure}

\vspace{20pt}

\begin{figure}[H]
\centering
\includegraphics[width=1.0\textwidth]{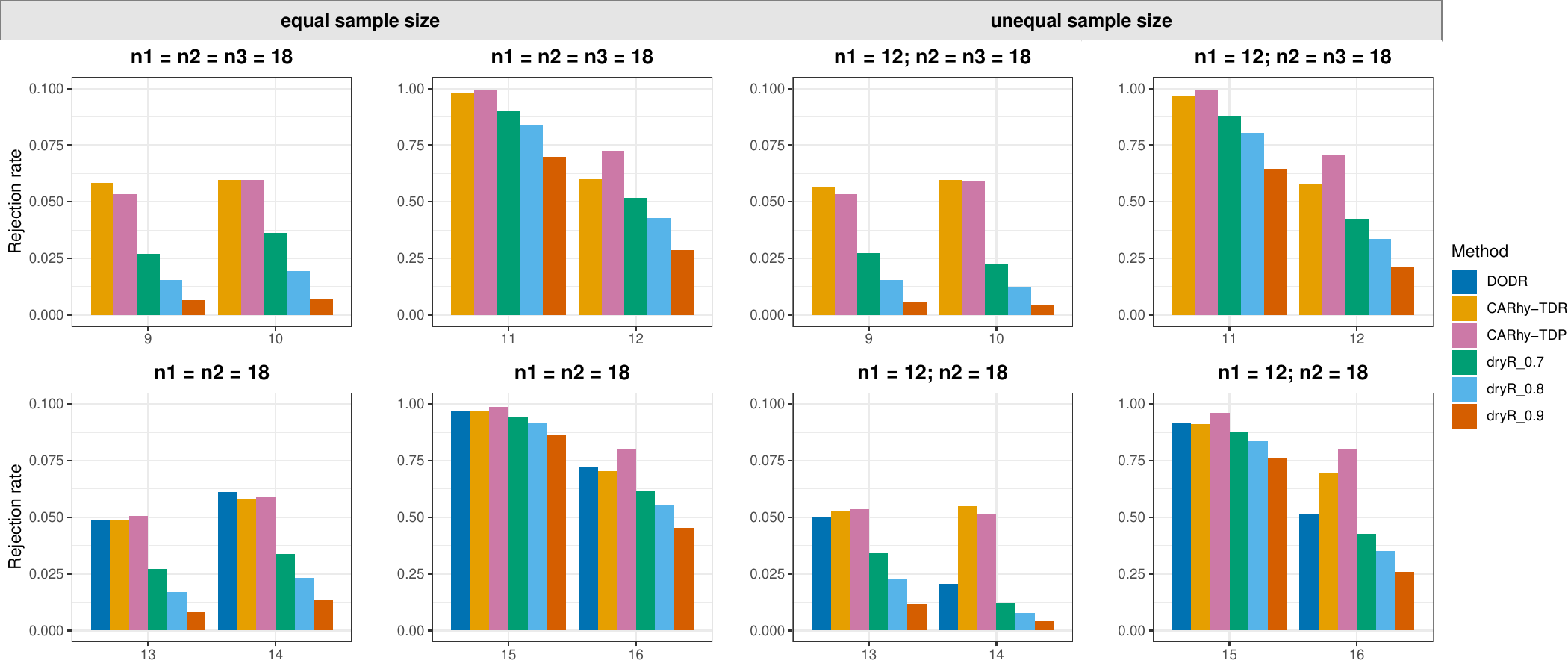}
\caption{Type-I error rate and power for testing differential phase (Table \ref{tab:S3}).}
\label{fig:S4}
\end{figure}

\begin{figure}[H]
\centering
\includegraphics[width=1.0\textwidth]{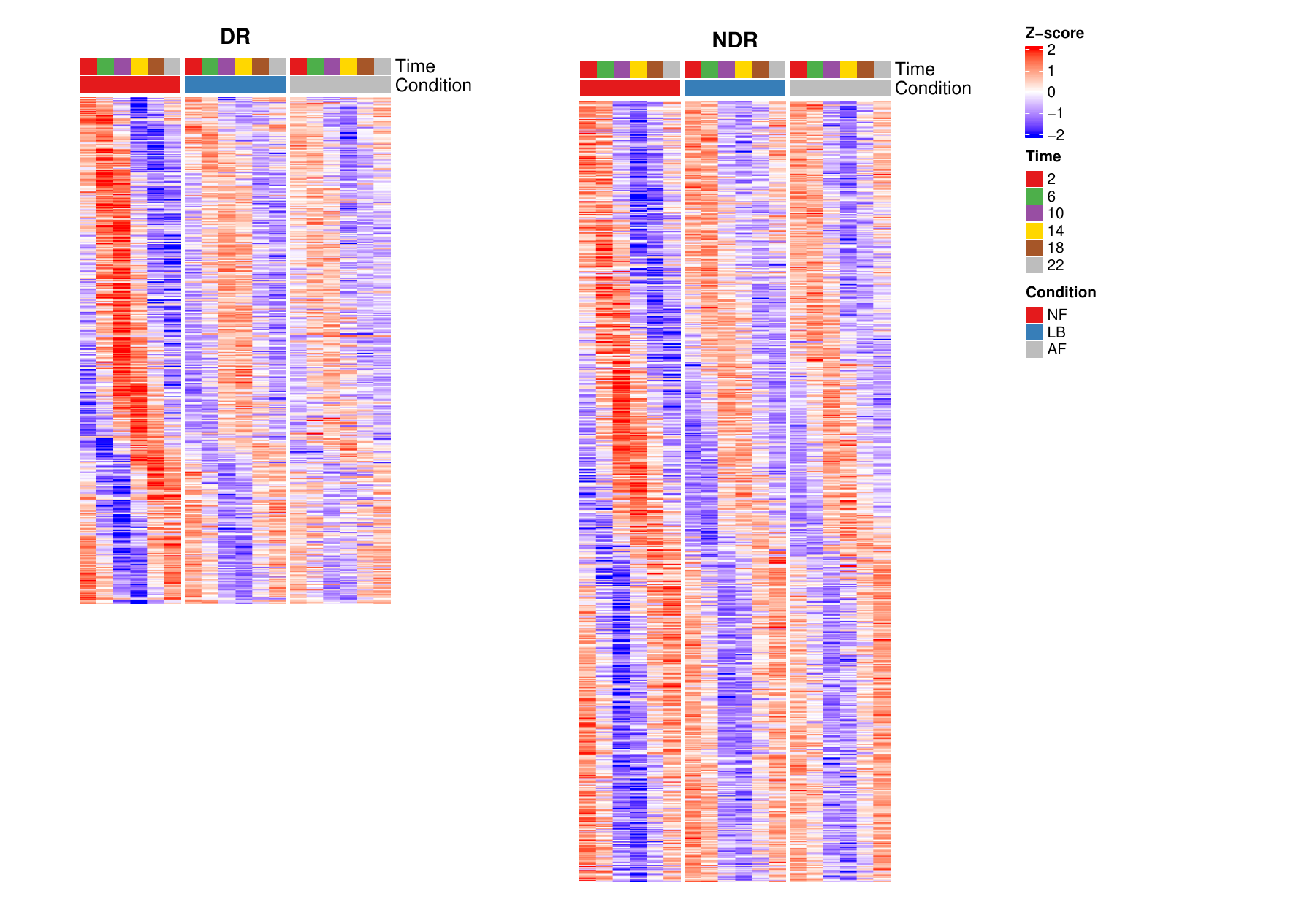}
\caption{Heatmap of differential rhythmic (DR) non-differential rhythmic (NDR) genes identified by CARhy-TDR.}
\label{fig:S8}
\end{figure}

\begin{figure}[H]
\centering
\includegraphics[width=1.0\textwidth,keepaspectratio]{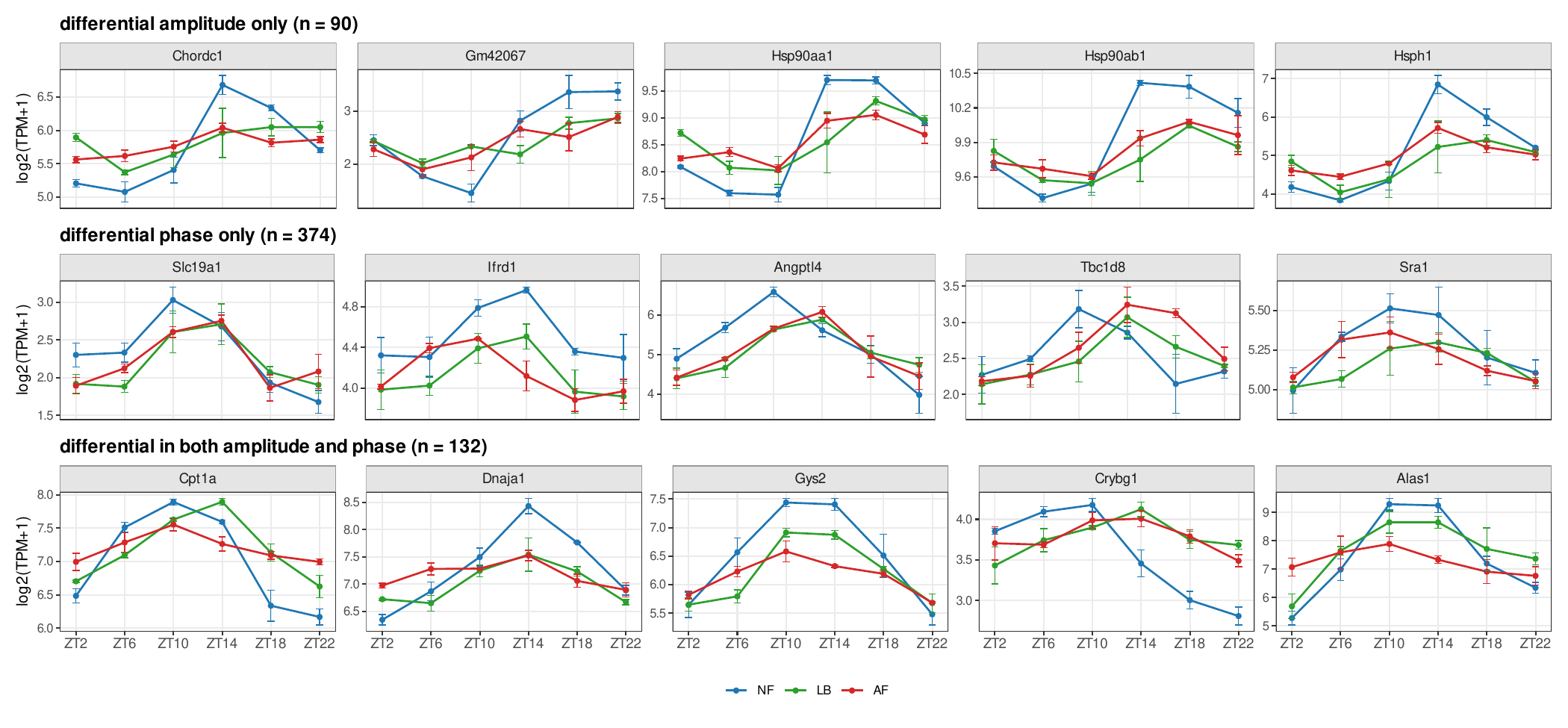}
\caption{CARhy-identified genes with differential amplitude only ($n = 90$), differential phase only ($n = 374$), or both differential amplitude and phase ($n = 132$) across NF, IB, and AF. The plots show representative genes and their expression patterns.}
\label{fig:S5}
\end{figure}

\begin{figure}[H]
\centering
\includegraphics[width=1.0\textwidth,keepaspectratio]{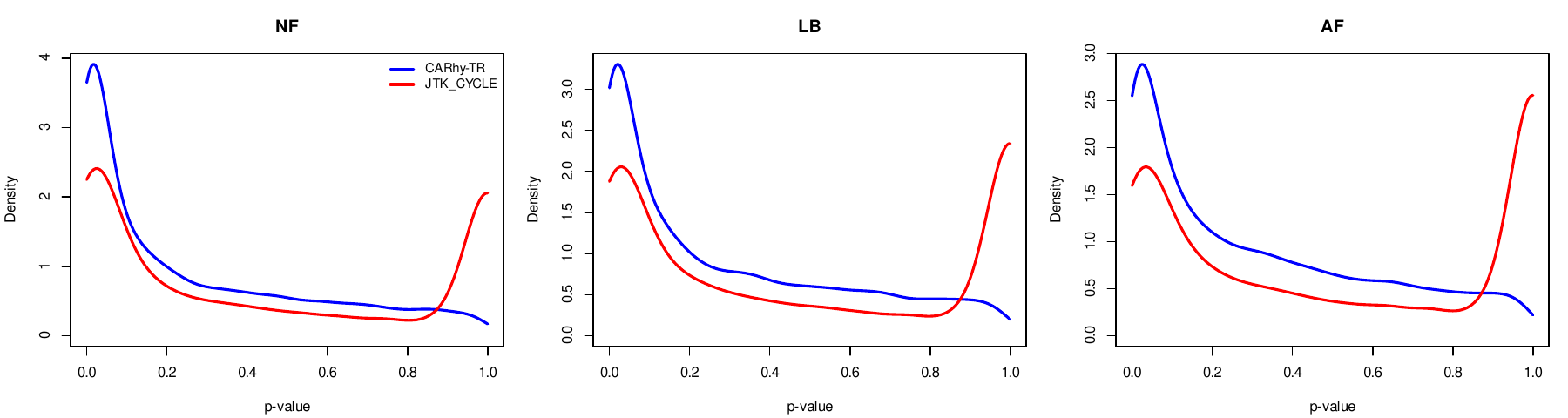}
\caption{P-value distributions for CARhy-TR and JTK\_CYCLE across conditions.}
\label{fig:S6}
\end{figure}

\begin{figure}[H]
\centering
\includegraphics[height=0.98\textheight,keepaspectratio]{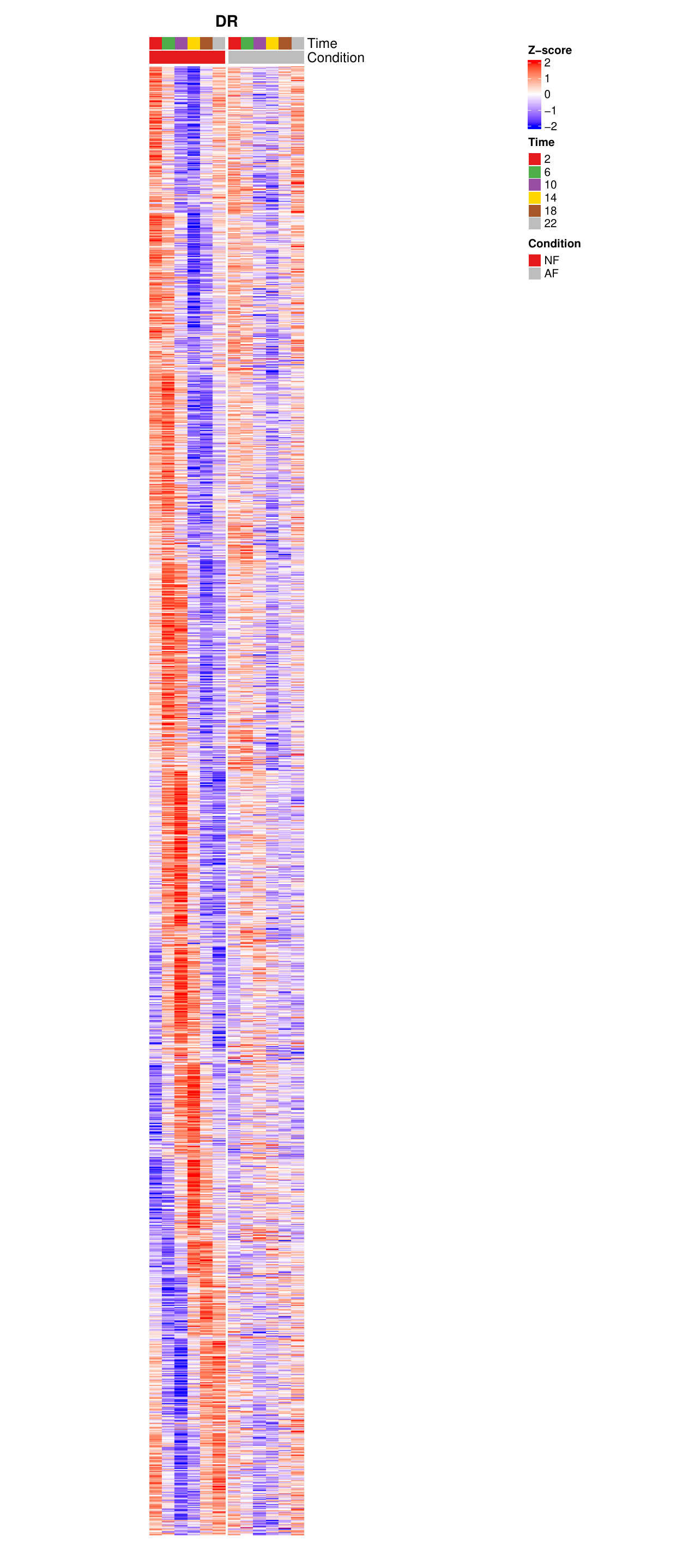}
\caption{Heatmap of differential rhythmic genes in reference set.}
\label{fig:S7}
\end{figure}

\end{document}